\documentclass[a4paper,twocolumns,english,usenatbib]{mn2e}

\usepackage{hyperref}

\usepackage[T1]{fontenc}
\usepackage[latin1]{inputenc}
\usepackage{times}

\usepackage{amsmath}
\usepackage{color}
\usepackage{graphicx}
\usepackage{amssymb}

\usepackage{xspace}
\usepackage{soul}
\definecolor{highlight}{rgb}{.70,.13,0.13}

\newcommand{\Me}{m_{\mathrm{e}}} 
\newcommand{\Qe}{q_{\mathrm{e}}} 
\newcommand{\Ne}{n_{\mathrm{e}}} 
\newcommand{\Nr}{n_{\mathrm{r}}} 
\newcommand{\mathi}{\mathrm{i}}
\newcommand{\mathe}{\mathrm{e}}
\newcommand{\Halpha}{H$_{\alpha}$\xspace}
\newcommand{\MAP}{\mathrm{MAP}}
\newcommand{\Prob}{\mathcal{P}}   
\newcommand{\Regul}{\mathcal{R}}  
\newcommand{\Cost}{\mathcal{Q}}   
\newcommand{\Lkl}{\mathcal{L}}    
\newcommand{\Imag}{\mathcal{I}\mathrm{m}} 
\newcommand{\Real}{\mathcal{R}\mathrm{e}} 
\makeatletter
\newcommand{\argmin}{\mathop{\operator@font arg\,min}\limits}
\newcommand{\argmax}{\mathop{\operator@font arg\,max}\limits}
\makeatother
\newcommand{\TFCmap}{\hat{\M{C}}_{\mathrm{MAP}}} 
\newcommand{\TFWmap}{\hat{\M{W}}_{\mathrm{MAP}}} 
\newcommand{\Cmap}{\M{C}_{\mathrm{MAP}}} 
\newcommand{\Wmap}{\M{W}_{\mathrm{MAP}}} 
\newcommand{\Cerr}{\M{C}_{\mathrm{n}}} 
\newcommand{\Serr}{\sigma_{\mathrm{n}}} 
\newcommand{\RegulWeight}{\mu_{\mathrm{s}}} 

\def\M#1{{\mathbf #1}}

\definecolor{Blue}{rgb}{0,0.08,0.65}
\definecolor{Red}{rgb}{0.65,0.08,0.05}
\definecolor{Green}{rgb}{0.15,0.45,0.25}
\definecolor{Purple}{rgb}{0.57,0.43,0.85}

\def\mathfrak#1{{\mathrm{#1}}}
\def\R#1{{\mathrm{#1}}}

\def\M#1{{\mathbf{#1}}}
\def\T#1{{{#1}^{\top}}}

      \def\be{{\bf e}}

\def\R#1{{\mathrm{#1}}}         

\def\be{\begin{equation}}
\def\ee{\end{equation}}
\def\ba{\begin{eqnarray}}
\def\ea{\end{eqnarray}}

\def\M#1{{\mathbf{#1}}} 
\def\T#1{{{#1}^{\top}}}         
\def\d{{\R{d}}}         

\usepackage{babel}
\makeatother

\begin{document}

\newcommand{\V}[1]{\mathbf{#1}}
\newcommand{\TR}{^{\mathrm{T}}}
\newcommand{\BF}[1]{\mathbf{#1}}
\newcommand{\RM}[1]{\mathrm{#1}}
\newcommand{\abs}[1]{|#1|}
\newcommand{\Abs}[1]{\left|#1\right|}
\newcommand{\norm}[1]{\Vert#1\Vert}
\newcommand{\Norm}[1]{\left\Vert #1\right\Vert }
\newcommand{\PDer}[2]{\frac{\partial #1}{\partial #2}}
\newcommand{\pder}[2]{\partial #1/\partial #2}

\title[Probing magnetic fields with multi-frequency polarized synchrotron
emission]{Probing magnetic fields in volume \\ with multi-frequency polarized synchrotron
emission.}

\author[{\Large Probing magnetic fields with multi-frequency polarized synchrotron
emission}]{J. Thi\'ebaut$^{1}$, S. Prunet$^{1}$\thanks{prunet@iap.fr}, C. Pichon$^{1,3}$ 
and E.~ Thi\'ebaut$^{2}$  \\
\textsl{\textcolor{black}{\normalsize $^{1}$Institut d'astrophysique
de Paris (UMR 7095), 98 bis boulevard Arago , 75014 Paris, France.}}\\
\textsl{\textcolor{black}{\normalsize $^{2}$Centre de Recherche Astronomique
de Lyon (UMR 5574), 9 avenue Charles Andr\'e, 69561 Saint Genis Laval
Cedex, France.}}\\
\textsl{\textcolor{black}{\normalsize $^{3}$Service d'Astrophysique, IRFU,  CEA-CNRS, L'orme des meurisiers, 91
470, Gif sur Yvette, France.}}}

\date{\today}

\maketitle

\begin{abstract}
We investigate the problem of probing the local spatial structure of the
magnetic field of the interstellar medium using multi-frequency polarized maps
of the synchrotron emission at radio wavelengths. 
We focus in this paper on the three-dimensional reconstruction of
the largest scales of the magnetic field, relying on the internal
depolarization (due to differential Faraday rotation) of the emitting
medium as a function of electromagnetic frequency. We argue that multi-band
spectroscopy in the radio wavelengths, developed
in the context  of high-redshift extragalactic HI lines,
can be a very useful probe of the 3D magnetic field structure of our
Galaxy when combined with a Maximum A Posteriori reconstruction technique. 

When starting from a fair approximation of the magnetic field, we are
able to recover the true one by using a linearized version of the corresponding
inverse problem.  The spectral analysis of this problem
allows us to specify the best sampling strategy in electromagnetic
frequency and predicts a spatially anisotropic distribution of
posterior errors. The reconstruction method is illustrated for reference fields extracted from realistic 
magneto-hydrodynamical simulations.
\end{abstract}


\section{introduction}
The problem of studying the magnetic field structure of our Galaxy using
measurements of the synchrotron emission of high energy electrons in the
Galactic magnetic field is an old one \citep{Ginzburg,Ruzmaikin,Beck}.  The
fact that the emitting medium is itself magnetized induces a differential
Faraday rotation of the different emission planes transverse to the line of
sight, resulting in a well known depolarization effect of the integrated
emission that depends strongly on the electromagnetic frequency. This effect,
described in the first place by \cite{Burn} in the case of a constant magnetic
field, has been further studied in semi-analytically for given functional
forms of the magnetic field; it has also been studied from the statistical
point of view in some asymptotic regimes \citep[see e.g.][]{Sokoloff}. In the
present work, we want to consider the more ambitious problem of using this
depolarization effect, together with the solenoidal character of the magnetic
field, to \emph{reconstruct} the magnetic field structure from a set of
polarized maps of the synchrotron emission of an ionized medium at different
electromagnetic frequencies.  With the upcoming prospect of detailed
Multi-band spectroscopy in the radio wavelengths \citep{HUBB,SKA}, developed
in the context of Galactic and high-redshift extragalactic HI lines, this type
of investigation should become possible.

A statistical inference of the measurement of the Galactic magnetic field
correlator as a function of scale from multi-frequency polarization
measurements has already been successfully achieved by \cite{Vogt} in the case
of the Faraday rotation of the polarized light from background objects by the
intra-cluster magnetized plasma. In this case, there is no depolarization
effect due to differential Faraday rotation, and the relationship between the
measured polarization at a given frequency and the polarization of light in
the source plane is linear in the (longitudinal) magnetic field strength. The
linearity of the problem makes the statistical analysis tractable in the
former case.  In the case that we investigate, the emitting and the rotating
medium are the same, which results in depolarization effects of the emitted
light. Moreover, the synchrotron emissivity itself depends non-linearly on the
field strength transverse to the line of sight. The reconstruction of the
magnetic field structure from the polarization data is in this case a
non-linear inverse problem.  Finally, we must note that to address the full
problem of reconstruction of the magnetic field from the depolarized
synchrotron emission we need in principle knowledge of both the thermal
electron spatial distribution $\Ne$ \emph{and} the spatial distribution of
cosmic ray electrons $\Nr$, when, in comparison, the inference of the magnetic
energy spectrum from the rotation measures of background sources only requires
knowledge of the thermal electron distribution.

In a first attempt at reconstructing the magnetic field, and for the
sake of clarity, we make the
assumption that the fluctuations of the thermal and cosmic ray electrons can
be neglected compared to the fluctuations in the magnetic field
itself. This assumption, if physically unrealistic, allows us to show
the specific influence of the magnetic field statistical properties on
the quality of the reconstruction. In
the first sections, we thus consider the electronic distributions (both thermal and
relativistic) as constant, and discuss the reconstruction of the magnetic
field using only the leading coupling coefficient in the equation of radiative
transfer.  In the (thin medium, strong rotativity) limit that we assume for
this work, this leading term is the usual Faraday term, responsible for the
rotation of the plane of polarization. We will assume that the Faraday
coefficient is dominated by the thermal electrons, which is a reasonable
assumption in non-relativistic astrophysical plasmas. 
Finally, in section~\ref{sec:valid}, we relax the unrealistic assumption of a
constant thermal electrons density, and show that our method can still
be used to reconstructed the magnetic field when the electronic
density is spatially varying but \emph{known a priori}, using
simulated data sets from a magneto-hydrodynamical (MHD) simulation. 

This paper is organized as follows: in section~\ref{sec:emission} we discuss
the fonctional dependence of the polarization of the synchrotron emission and
its variation with electro-magnetic frequency on the underlying magnetic
field. We present a discretized version of this functional dependence that
will be useful in the context of the reconstruction from discrete polarization
data. In section~\ref{sec:linn} we investigate the reconstruction of the
magnetic field from simulated multi-frequency polarized data, when the
functional dependence on the magnetic field has been linearized around a
"mean" field. Taking advantage of the linear nature of this approximate
problem, we give a strategy for choosing the best electromagnetic frequencies
of observation, and investigate the statistical anisotropy of the magnetic
field reconstruction errors.  Finally, in section~\ref{sec:valid}, we
investigate the validity of the linearization procedure used in the precedent
section, as a function of the quality of our prior knowledge of the magnetic
field structure. We show how the approximate, linearized inverse problem
investigated in this work could be used as a building block of the fully
non-linear reconstruction problem.  We emphasize that any gradient-based
non-linear minimization algorithm can be decomposed into linear sub-problems, thus
justifying the study of the linearized problem.  In this context, we
investigate how the conditioning of the linearized problem varies with the
properties of the reference magnetic field around which the problem is being
linearized. In particular it is illustrated on a realistic
reference field from a MHD simulation.  Finally, using the same MHD
simulation data, we show that our method can deal with a non-constant
electronic density, provided it is known a priori.
In section~\ref{sec:Conclusion}, we summarize the main results of the paper,
recalling the main simplifying assumptions used to derive them (notably the
 assumed-known electronic density hypothesis) and discuss how this assumption
could be possibly alleviated by additional data (e.g., \Halpha,
free-free) or by using second-order coupling terms involving the
circular polarization in the case of relativistic sources
(see~\ref{sec:Circular-Polarization}).  
We conclude on how the different results of the paper could be used to tackle the
fully non-linear reconstruction of the magnetic field.

\section{Polarized emission} \label{sec:emission}

Our objective is to recover the magnetic field given observed
  polarization maps at different wavelengths.  We tackle this ill-posed
  problem by means of an inverse problem approach \citep{Tarantola} which
  involves recovering the magnetic field that gives a
  polarization consistent with the observations while obeying some a priori
  properties.  These priors are strict constraints, such as $\nabla\cdot B=0$,
  to insure that the sought field is physically meaningful and a   regularization
  to lever the degeneracies of the inverse problem while avoiding artifacts
  due to noise amplification.  We first derive the direct model of the
  polarization given the magnetic field and then introduce the inverse problem
  approach in a Bayesian framework.

\subsection{Direct model} \label{sec:form}

We only consider here the Faraday rotation in the transfer equation,
and neglect all other coupling terms. In this case, the transfer equation of
the Stokes parameters of linear polarization $(Q,U)$ can be integrated
formally. We assume here that the density of electrons is constant, or that
its fluctuations are only important on scales that are not considered here.

Consider a slab of ionized magnetized medium of width $L$ which is emitting
synchrotron radiation.  The polarized emission, as a function of frequency,
integrated over the line of sight then reads \citep{Sokoloff}:
\begin{equation}
  P \equiv Q + \mathi\,U
  = \int\epsilon(\V{r})\,\mathe^{2\,\mathi\,\psi(\V{r})}\d z\,\label{eq:PQU},
\end{equation}
with $Q$ and $U$ are the usual Stokes parameters, $\epsilon(\V{r})$ the
synchronton emissivity which obeys:
\begin{equation}
  \epsilon(\V{r}) = A \,\Nr(\V{r})\, |B_{\perp}(\V{r})|^{\frac{\gamma+1}{2}}
  \, \nu^{-\frac{\gamma-1}{2}} \, ,
  \label{eq:defeps}
\end{equation}
and $\psi(\V{r})$ the sum of the Faraday rotation and the primordial
orientation:
\begin{equation}
  \psi(\V{r}) = \pi/2 + \arctan\left(B_{y}/B_{x}\right)
  + \frac{K}{\nu^{2}} \, \int_{z}^{0} \,\Ne B_{z} \, \d z'\,,
  \label{eq:defpsi}
\end{equation}
where $\V{r}\equiv(x,y,z)=(\V{x}_{\perp},z)$ is the coordinate in the slab,
$\nu$ is the frequency, and $B=(B_x,B_y,B_z)=(B_\perp,B_z)$ is the magnetic
field.  In equation~(\ref{eq:defpsi}), $K$ reads:
\begin{equation}
  K = \frac{\Qe^3}{8\,\pi^2\,\Me^2\,c\,\epsilon_0}\,, 
\end{equation}
while, in equation~(\ref{eq:defeps}), $A$ is given by
\begin{equation}
  A = \frac{\sqrt{3}\,E_0^{\gamma}\,\Qe^3}{16\,\pi\,\epsilon_0\,\Me\,c}\left(\frac{3\,\Qe}{2\,\pi\,\Me^3\,c^4}\right)^{\!\!\!\frac{\gamma-1}{2}}\Gamma\!\left(\!\frac{3\,\gamma-1}{12}\!\right)\Gamma\!\left(\!\frac{3\,\gamma+1}{12}\!\right) \,,\nonumber
\end{equation}
where $E_0$ is the energy scale of the relativistic electron spectrum, $\Me$
and $\Qe$ stand for the mass and the charge of the electron, $\Ne$ and $\Nr$
are the thermal and relativistic electron densities supposed constant, while
the exponent $\gamma$ stands for the spectral index of the cosmic ray
electrons, $c$ is the speed of light, $\epsilon_0$ is the electric
permittivity and $\Gamma$ is the Euler gamma function.  The lengths are in
kilo-parsec (kpc) and so the density in kpc${}^{-3}$, the magnetic fields in
micro-Gauss ($\mu$G) and the frequencies in giga-Hertz (GHz).  Re-expressing
the intrinsic polarization phase in terms of powers of the magnetic field
components, we get the following expression for the polarization:
\begin{align}
  P(\V{x}_{\perp},\nu)
  &= A\,\nu^{-\frac{\gamma - 1}{2}} \int_{-\infty}^{0}
  \,\Nr(\V{x}_{\perp},z)\left(B_{x}^{2} + B_{y}^{2}\right)^{\frac{\gamma-3}{4}}(\V{x}_{\perp},z)
  \notag \\
  &\quad{\times}\:
  \left(B_{x}^{2} - B_{y}^{2} + 2\,\mathi\,B_{x}\,B_{y}\right)(\V{x}_{\perp},z)
  \notag \\
  &\quad{\times}\:
  \exp\!\left(
    \frac{2\,\mathi\,K}{\nu^{2}}
    \int_{z}^{0} \,(\Ne B_{z})(\V{x}_{\perp},z'') \, \d z''
  \right) \, \d z \, .
  \label{eq:defP}
\end{align}
As real data come in discrete form, let us discretize this
expression by replacing all integrals with sums, assuming a regular
discretization grid that will be defined more precisely
below. Equation~(\ref{eq:defP}) then reads
\begin{align}
  P(\V{x}_{\perp},\nu)
  & = A \, h \, \nu^{-\frac{\gamma-1}{2}} \sum_{z}
  \,\Nr(\V{x}_{\perp},z)\left(B_{x}^{2} + B_{y}^{2}\right)^{\frac{\gamma-3}{4}}(\V{x}_{\perp},z)
  \notag \\
  & \quad{\times}\:
  \left(B_{x}^{2} - B_{y}^{2} + 2\,\mathi\,B_{x}\,B_{y}\right)(\V{x}_{\perp},z)
  \notag \\
  & \quad{\times}\:
  \exp\!\left(
    \frac{2\,\mathi\,K\,h}{\nu^{2}}
    \sum_{z'}\theta_{\rm H}(z'-z)\,(\Ne\,B_{z})(\V{x}_{\perp},z')
  \right) \, .
  \label{eq:defPdisc}
\end{align}
Here $\theta_H$ is the Heaviside function ($\theta_H(x)=1$ for $x\ge 0$ and 0
elsewhere), and $h$ the discretization length along
$z$. Equation~(\ref{eq:defPdisc}) is formally a function of $\V{B} \equiv
\left\{ (B_x,B_y,B_z) \right\}_{\V{r}}$ where we use bold symbols to represent
the discretized vector fields and $\V{r}$ is a triple index spanning the
magnetized volume on a regular cubic mesh with cell size $h$.

The solution to the inverse problem will be obtained by means of minimization
of some merit function (as explained in what follows), we therefore need to
compute the partial derivatives of the polarization with respect to the
magnetic field.  Let us first compute the derivatives with respect to the
transverse components of the field:
\begin{align}
  \frac{\partial P(\V{x}_{\perp},\nu)}{\partial B_{x}(\V{r}')}
  &= 
  \delta_\mathrm{D}(\V{r} - \V{r}')
  \, A \,\Nr(\V{r}')\, h \, \nu^{-\frac{\gamma-1}{2}}
  \, (B_{x}^{2} + B_{y}^{2})^{\frac{\gamma-7}{4}}(\V{r}')
  \notag \\
  &\hspace{-13mm}{\times}\left[\!
    \frac{1 + \gamma}{2} \, B_{x}^{3} +
    \frac{7 - \gamma}{2} \, B_{y}^{2} \, B_{x} +
    \mathi\,(\gamma - 1) \, B_{x}^{2}\, B_{y} +
    2\,\mathi\,B_{y}^{3}
  \!\right]\!\!(\V{r}')
  \notag \\
  &\hspace{-13mm}{\times}\:
  \exp\!\left(\frac{2\,\mathi\,K\,h}{\nu^{2}}
    \sum_{z''}\theta_{\rm H}(z''-z')\,(\Ne\,B_{z})(\V{x}'_{\perp},z'')\right)\,, 
  \label{eq:defdPdBx}
\end{align}
with $\V{r}=(\V{x}_{\perp},z)$, $\V{r}'=(\V{x}'_{\perp},z')$ and
$\delta_{\mathrm{D}}$ Dirac's delta function.  The derivative with respect to
$B_{y}$ follows closely, with the square bracket term becoming:
\begin{equation}
  \left[
    - \frac{1 + \gamma}{2}\,B_{y}^{3}
    - \frac{7 - \gamma}{2}\,B_{x}^{2}\,B_{y}
    + \mathi\,(\gamma - 1)\,B_{x}\,B_{y}^{2}
    + 2\,\mathi\,B_{x}^{3}
  \right]
  \label{eq:defdPdBy}
\end{equation}
which corresponds to a $\pi/2$ rotation in the plane perpendicular to the LOS. 
We see that in both cases the phase term is unaffected since it is
only a function of the longitudinal magnetic field component $B_{z}$.
Finally let us compute the derivative with respect to $B_{z}$:
\begin{eqnarray}
\nonumber
\frac{\partial P(\V{x}_{\perp},\nu)}{\partial B_{z}({\bf \V{r}'})} &=&
\delta_{\rm D}(\V{x}_{\perp}-\V{x}'_{\perp})2iKAh^{2}\nu^{-\frac{\gamma+3}{2}}
\\&& \hskip -2cm
\nonumber 
\times\sum_{z}\,\Nr(\V{x}'_{\perp},z)(B_{x}^{2}+B_{y}^{2})^{\frac{\gamma-3}{4}}(B_{x}^{2}-B_{y}^{2}+2iB_{x}B_{y})(\V{x}'_{\perp},z)
\\ &&  \hskip -2.5cm
\times \theta_{\rm
  H}(z'-z)\exp\left({\frac{2iKh}{\nu^{2}}\sum_{z''}\theta_{\rm H}(z''-z)(\,\Ne B_{z})(\V{x}_{\perp},z'')}\right)\,. 
\label{eq:defdPdBz}
\end{eqnarray}
We note that here the phase term, not the emissivity layer term, is involved.
The case $\gamma=3$ is detailed in Appendix~\ref{sec:gamma3} and leads to a
simplification of the above equations.

\subsection{Maximum A Posteriori formulation}
\label{sec:MAP}

From the direct model, we can express the observed data as:
\begin{equation}
  \label{eq:data}
  d_m = P\bigl((\V{x}_{\perp},\nu)_m,\V{B}\bigr) + e_m \, ,
\end{equation}
with $m$ an index which spans the mixed frequency position-on-the-sky cube,
$(\V{x}_{\perp},\nu)_m$ the corresponding coordinates, $\V{B}$ the actual
magnetic field and $e_m$ an error term which accounts for noise and model
approximations.  Using vector notation, equation~(\ref{eq:data}) simplifies
to: $\V{d} = \V{P}(\V{B}) + \V{e}$ with $\V{d}=\{ d_m\}$ the vector collecting
all the observations,
$\V{P}(\V{B})=\{P\bigl((\V{x}_{\perp},\nu)_m,\V{B}\bigr)\}$ and
$\V{e}=\{e_m\}$.  Our inverse problem is to recover the magnetic field vector,
$\V{B}$, given some noisy measurements of the polarization, $\V{d}$.  Due to
the unknown errors in equation~(\ref{eq:data}) and to possible strict
degeneracies of the direct model, there is not a unique magnetic field that
yields a polarization consistent with the observations.  We therefore need
some means to select a unique solution and, hopefully, the \emph{best} one
given the data.


Probabilities provide a consistent framework to define such a solution; we
thus define the sought magnetic field as being the most likely given the
observations.  It is the one which maximizes the posterior probability:
\begin{equation}
  \label{eq:mapdef}
  \V{B}_\MAP = \argmax_\V{B}\Prob(\V{B}|\V{d}) \, ,
\end{equation}
and which is termed as the \emph{maximum a posteriori} (MAP) solution
\citep[see e.g.][]{PT98}.  By Bayes' theorem,
$\Prob(\V{B}|\V{d})=\Prob(\V{d}|\V{B})\,\Prob(\V{B})/\Prob(\V{d})$, and since
$\Prob(\V{d})$ does not depend on the sought parameters $\V{B}$, this amounts
to maximizing $\Prob(\V{d}|\V{B})\,\Prob(\V{B})$.  The term $\Prob(\V{d}|\V{B})$
is the likelihood of the data given the model, while the term $\Prob(\V{B})$
accounts for any \emph{a priori} knowledge about the magnetic field.  We can
anticipate two types of priors: (i) the strict constraint that, to be
physically meaningful, the field should be solenoidal: $\nabla\!B=0$; (ii)
some so-called \emph{regularization} constraint to overcome the
ill-conditioning of the inverse problem and to enforce the unicity of the
solution.  Without loss of generality, we state that the probabilities writes:
\begin{align}
  \label{eq:likelihood}
  \Prob(\V{d}|\V{B})
  &= \kappa_1\,\exp\!\left(-{\textstyle\frac{1}{2}}\,\Lkl(\V{B})\right)\,, \\
  \Prob(\V{B}) &=\left\{
  \begin{array}{ll}
    \kappa_2\,\exp\!\left(-{\textstyle\frac{1}{2}}\,\Regul(\V{B};\mu)\right)\,,
    &\text{if \,\,\,\,$\nabla\!B=0$,}\\[1ex]
    0 &\text{otherwise.}\\
  \end{array}\right.
\end{align}
where the factors $\kappa_1$ and $\kappa_2$ do not depend on $\V{B}$ and $\mu$ accounts
for parameters to tune the regularization.  Finally, taking the
log-probabilities and discarding constants, the maximum a posteriori magnetic
field writes:
\begin{align}
  \label{eq:mapsol}
  \V{B}_\MAP = \argmin_{\V{B}, \nabla\!B=0} \Cost(\V{B}) \, ,
  \intertext{with:}
  \Cost(\V{B}) = \Lkl(\V{B}) + \Regul(\V{B};\mu) \, ,
\end{align}
which is the objective function.  Before going into the details of the
expressions of $\Lkl(\V{B})$ and $\Regul(\V{B};\mu)$ we can already note that
the solution $\V{B}_\MAP$ will depend on the data $\V{d}$ and on the
regularization parameters $\mu$.  The value of $\mu$ can be chosen, e.g., to
provide the best bias-variance compromise on the sought solution
\citep{wahba,golub}.

\subsubsection{Likelihood}

Assuming Gaussian statistics for the noise and model errors, the likelihood of
the data is the so-called $\chi^2$ and writes:
\begin{equation}
  \label{eq:defchi2}
  \Lkl(\V{B})
  = \T{\bigl(\V{d} - \V{P}(\V{B})\bigr)}\cdot\Cerr^{-1}
  \cdot\bigl(\V{d} - \V{P}(\V{B})\bigr)
\end{equation}
with $\Cerr$ the covariance matrix of the errors.  There is a slight
issue here because we are dealing with complex values.  Since complex numbers
are just pairs of reals, complex valued vectors such as $\V{d}$,
$\V{P}(\V{B})$ and $\V{e}$ can be \emph{flattened} into ordinary real vectors
(with doubled size) to use standard linear algebra notation.  This is what is
assumed in equation~(\ref{eq:defchi2}).  Under these conventions, the
covariance matrix of the errors writes
$\Cerr=\langle\V{e}\cdot\T{\V{e}}\rangle$ with $\T{}$ to denote
transposition.

\subsubsection{Regularization}\label{sec:penalty}

The regularization term $\Regul(\V{B};\mu)$ implements loose constraints to
avoid over-fitting the data and enforce local unicity of the solution (see
section~\ref{sec:NL}).  Requiring that the magnetic field be as smooth as
possible (while being consistent with the data) matches these requirements and
is supported by physics since the magnetic field should have no
discontinuities.  To simplify further computations, we choose the following
particular expression of the regularization $\Regul$ to favor the smoothness
of the field:
\begin{equation}
  \mathcal{R}
  = \RegulWeight\|\Delta^{\alpha/4}\V{B}\|^{2}
  \propto \RegulWeight \, \sum_\V{k} |\V{k}|^\alpha |\hat{\V{B}}|^2
  \,,
  \label{eq:divb2}
\end{equation}
which scales as the integrated norm of the spatial Laplacian of the field to
the power $\alpha/4$.  For a periodic field, this generic smoothing penalty is
diagonal in Fourier space.  In addition, if the model $\V B$ is Gaussian and
scale invariant, then $\alpha$ may be chosen to be the power law index of the
power spectrum $|\hat{\V{B}}|_k^2$ of the field.  In this case, choosing the
specific value of the hyperparameter, $\RegulWeight=1/|\hat{\V{B}}|_{k=1}^2$,
the MAP solution correspond to the minimal variance Wiener filtered data.

\subsubsection{Imposing $\nabla\!B=0$}
\label{sec:div}

For simplicity, we assume here that the magnetic field is multi-periodic, with
period $L$ in all three directions.  We may then rewrite the magnetic field
as:
\begin{equation}
  \V{B}=\V{F}^{-1}\cdot
  (\hat{B}_{\perp1}\V{e}_{\perp1}+\hat{B}_{\perp2}\V{e}_{\perp2})\equiv 
  \M{\Pi} \cdot \V{B}\,, \label{eq:defproj}
\end{equation}
where {$\V{F}^{-1}=\V F^{\dagger}/N_{\mathrm{r}}^3$} and $\V{F}$ is the
forward DFT operator, ($\V{e}_{\parallel}\equiv\V{k}/\abs{\V{k}}$,
$\V{e}_{\perp1},\,\V{e}_{\perp2}$) form a spherical basis in Fourier space,
while $\hat{B}_{\perp,i}$, i=1,2 are the projections over that basis of the
Fourier component, $ \V{\hat B}\equiv \V{F}\cdot \V{B}$ of the field.
Equation~(\ref{eq:defproj}) defines the projector $\M{\Pi}=\M{F}^{-1}\cdot
(\V{e}_\perp \otimes \V{e}_\perp) \cdot \M{F}$.  Such a field satisfies by
construction
\begin{equation}
  {\bf k}\cdot\hat{\V{B}}\equiv0\,,\quad{\rm which\
 implies}\quad\V{\nabla}\cdot B\equiv0\,.\label{eq:divb1}
\end{equation}
In fact, there is a slight complication at the Nyquist frequencies where only
one component of the field is free, see appendix \ref{app:Bfields}.

Note that the divergence free condition could also be imposed by other means
\citep[see e.g.][]{Nocedal_Wright-2006-numerical_optimization}.  For instance,
by adding a quadratic penalty term like $\sum_{\V{r}}(\nabla B)_{\V{r}}^2$ to
the total penalty $\Cost(\V B)$.  We however found that, in practice, the
projector $\M{\Pi}$ led to a better conditioned reconstruction problem.

\begin{figure*}
\includegraphics[scale=0.55]{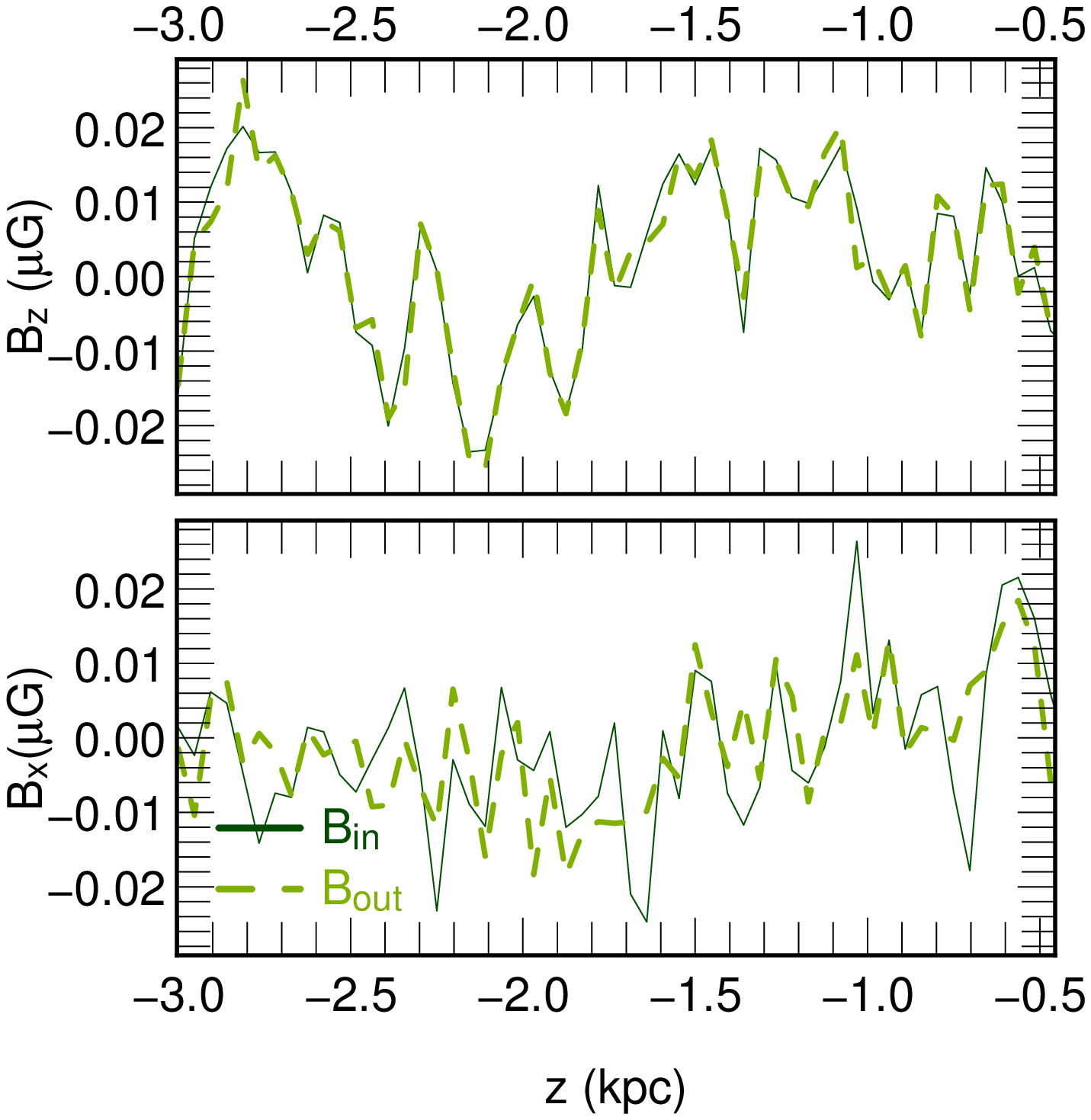}
\includegraphics[scale=0.55]{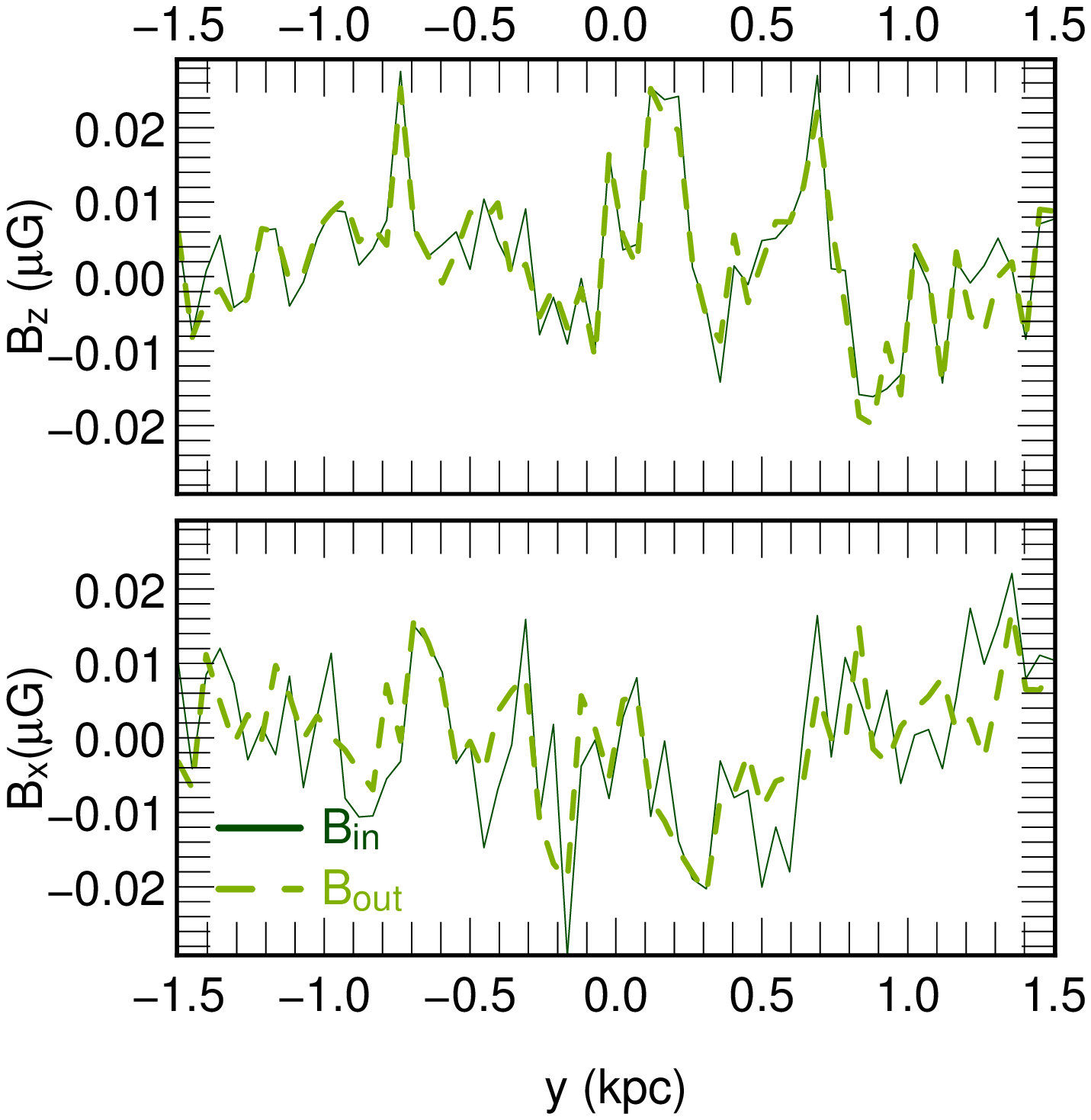}
\includegraphics[scale=0.5]{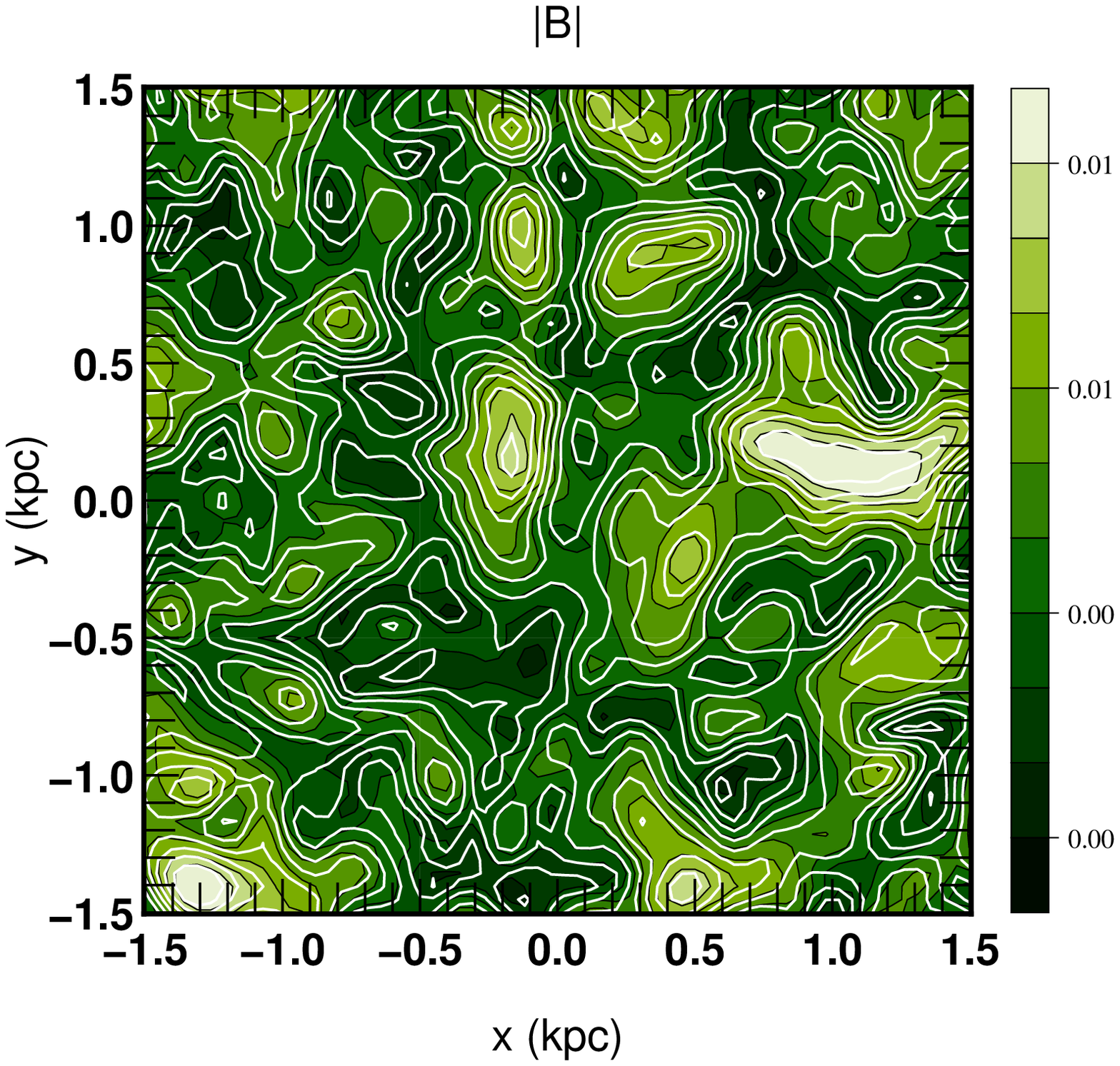}
\includegraphics[scale=0.5]{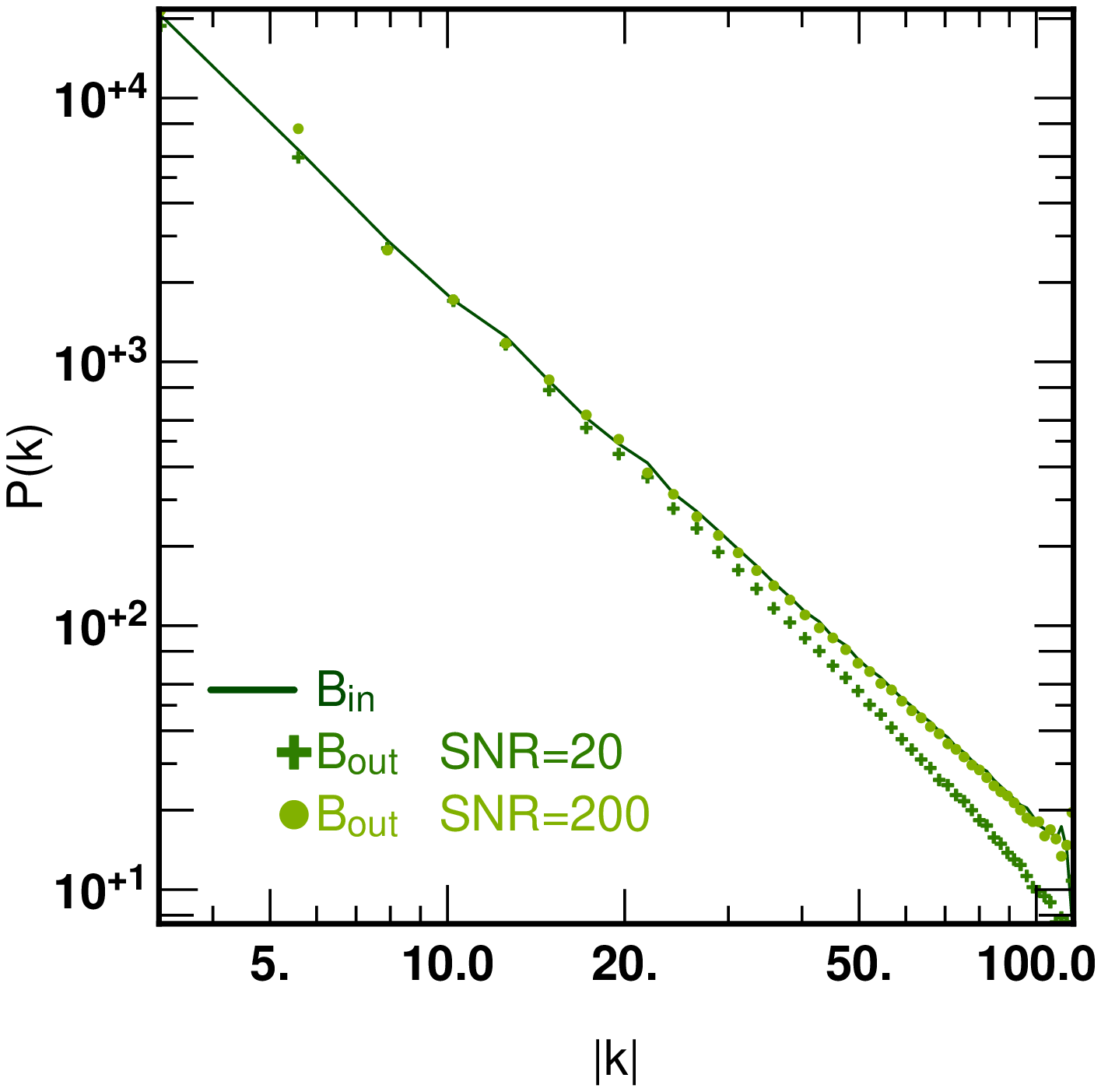}
\caption{\emph{Top:} input (\emph{solid lines}) and recovered (\emph{dashed
    lines}) $x$ and $z$ components of the field along a LOS (\emph{left}) and
  along the $y$ transverse direction (\emph{right}). The $y$ component and the
  $x$ direction are not plotted since very close to the $x$ component and the
  $y$ direction. One can see that the $z$ component is better reconstructed
  than the $x$ or $y$ components which is consistant with the variance
  measurements and the global conditioning of the problem (see section
  \ref{sec:var_SNR}). The reconstruction is carried on a $N_{\V r}=64$ grid
  with $N_{\nu}=64$ frequency channel. The data are generated linearly (see
  section \ref{sec:linn}) with a SNR$=20$. \emph{Bottom left:} maps of
  $\abs{B}$ for a transverse section after smoothing of the fields. The green
  images represents the input field while the superposed white contours show
  the recovered one. \emph{Bottom right:} power spectra of the input field
  (\emph{solid line}) and the recovered one (\emph{crosses}).  As expected,
  the recovered power spectrum is damped at higher frequencies because of the
  regularization. To illustrate this we added the power spectrum of a
  reconstruction with SNR=200.
   \label{fig:64cube}}
\end{figure*}

\subsection{Implementation}
\label{sec:mpl}

Given equations~(\ref{eq:defchi2}) and (\ref{eq:divb2}) the objective function
writes:
\begin{equation}
  \Cost = (\V{P}-\V{d})^{\dagger}\cdot\Cerr^{-1} \cdot (\V{P}-\V{d})
  + \RegulWeight\,\|\Delta^{{\alpha}/{4}}\V{B}\|^{2}\,.
  \label{eq:defQ}
\end{equation}
To minimize $\Cost(\V{B})$, we used a variable metric limited memory
optimization method with BFGS updates \citep{Nocedal-1980-vmlm} called VMLM
and implemented in OptimPack\footnote{OptimPack is freely available
  at\ {\small\texttt{http://www-obs.univ\-lyon1.fr/labo/perso/eric.thiebaut/optimpack.html}}.}
\citep{optimpack}.  Finding the optimal solution, equation~(\ref{eq:mapsol}),
involves computing the gradient of equation~(\ref{eq:defQ}) with respect to
$\V B$.  Now differentiating equation~(\ref{eq:defchi2}) with respect to a
magnetic field components we get
\begin{equation}
  \frac{\partial\chi^{2}}{\partial \V{B}_{i}}
  = 2\,\Real\left[(\V{P}-\V{d})^{\dagger}\cdot\Cerr^{-1}\cdot
    \frac{\partial \V{P}}{\partial \V{B}_{i}}\right]\,,
\end{equation}
where ${\partial \V{P}}/{\partial \V{B}_{i}}$ for $i=x,y,z$ are given by
equations~(\ref{eq:defdPdBx}) and (\ref{eq:defdPdBz}).  Similarly,
differentiating equation~(\ref{eq:divb2}) with respect to $\V B$ yields
\begin{equation}
  \frac{\partial \Regul}{\partial \V{B}_{i}}
  = \RegulWeight \, \M{F}^{-1}\cdot \V{\hat B}_i |\V{k}|^{\alpha}  \,.
\end{equation}
The VMLM algorithm is a quasi-Newton method which proceeds by solving
successive linear problems.  Let us therefore first consider in the next section
a linearized version of our inverse problem, which may correspond to a
physically motivated problem when a good first guess for the magnetic field is
known.

Note finally that equations~(\ref{eq:defdPdBx}) and (\ref{eq:defdPdBz}) imply
that ${\partial\chi^{2}}/{\partial \V{B}_{i}}=\mathbf 0$ at $B_x=B_y=0$.  Note
also that if $(B_x,B_y,B_z)$ is a solution to equation~(\ref{eq:defP}), so is
$(-B_x,-B_y,B_z)$.  Consequently we expect that the $\chi^2$ will be strongly
multivalued as a function of $\V B$\footnote{ For instance a magnetic loop
  close to the $z$ axis (where $B_x$ and $B_y\sim0$) and its mirror image by
  symmetry along the $z$ axis have the same $\chi^2$ and almost zero gradient.
}. The smoothing penalty should in part prevent a pixel-by-pixel flip of the
$x$ and $y$ component.  It remains nonetheless to be shown that the zero
divergence condition is sufficient to avoid flipping the field in regions
bound by zeros of these two components, if such regions exist.  Addressing these issues will be the
topic of another paper.

\section{Linearization} \label{sec:linn}
Let us first consider the situation when a fairly good guess for
the overall magnetic field, $\V{B}_{0}$, is known, on the basis, say
of a first large scale investigation, or via some modelling of the
field as a function of the underlying density \citep[e.g.][]{CDYZmodel,GMFmodels}. Let us then seek the
departure from this guess. It is then legitimate to assume $\V{B}=\V{B}_{0}+\delta\V{B}$,
with, possibly (if the prime guess is accurate enough)
$\delta\V{B}/\abs{\V{B}_{0}}\ll1$, so that equation (\ref{eq:defP}) becomes:
\begin{equation}
\delta \V{P}\equiv\left(\frac{\partial \V{P}}{\partial\V{B}}\right)_{\V{B}_{0}}\cdot\,\delta\V{B},\label{eq:defPlin}\end{equation}
where the tensor $\partial \V{P}/\partial \V{B}_{i}$ is given by its components,
equations (\ref{eq:defdPdBx}), (\ref{eq:defdPdBy}) and (\ref{eq:defdPdBz}), while $\delta \V{P}\equiv \V{P}-\V{P}(\V{B}_{0})$.
Now equation (\ref{eq:defPlin}) is  likely to be a much better behaved equation
as the linearity warrants convexity of the objective function, hence  the formal unicity of the solution. 

In this paper, we will address two linear problems in turn, one of academic interest, to understand 
the properties of the inverse problem at hand, while the second one  should allow us
 to carry realistic reconstructions, in the regime when a fair reference field is known.  
Specifically, we will first assume that the (noise free) data is in the image of $\left({\partial
 \V{P}}/{\partial\V{B}}\right)_{\V{B}_{0}} $:
\begin{equation}
\V{d}\equiv \delta \V{P}_{L}=\left({\partial
 \V{P}}/{\partial\V{B}}\right)_{\V{B}_{0}}\cdot\,\delta\V{B}+\V{e}\,,
\quad \mbox{linear problem (I),} \nonumber
 \end{equation}
 while for the second problem (the so called Gauss-Newton approximation)
\begin{equation}
\V{d}\equiv \delta \V{P}_{\rm PL}\equiv \V{P}-\V{P}(\V{B}_{0})+\V{e}\,,
\quad \mbox{pseudo linear problem (II).} \nonumber
\end{equation}
We investigate the linear problem in this section and the pseudo-linear problem in section~\ref{sec:valid}.

\subsection{Linear reconstruction}\label{sec:rec}

\begin{figure}
\includegraphics[angle=90,scale=0.35]{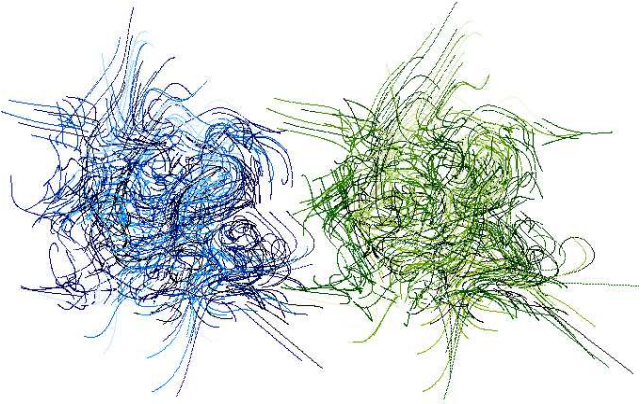}
\caption{Field lines of the input (\emph{left}) and the recovered
  (\emph{right}) fields for a $64^3$ reconstruction with $N_{\nu}=64$
  frequencies. 
  The fields correspond to a reconstruction with a SNR of 200.
 \label{fig:field_lines}}
\end{figure}

Let us illustrate our method on a
problem of realistic scales. This first simulation is carried on a $N_\V{r}^3$ grid ($N_\V{r}=64$) with
$N_\nu=64$ frequencies. The reference field $\V{B}_{0}$ is chosen constant and
set to $1\,\mu$G  everywhere for each component, the power spectrum of the
perturbation field $\delta\V{B}$ has a power law index $\alpha=2$ and its RMS
is $0.01$. 
Data are
simulated  linearly (see section \ref{sec:linn}) and are noised with a
SNR$=20$.
Figure \ref{fig:64cube} illustrates the quality of the reconstruction.
The \emph{top panel} represents the $x$ and $z$ components along the LOS ($z$
direction) or transverse ($y$ direction) for a given pixel. As the results for
the $y$ component and the $x$ direction are similar to the $x$ component and
the $y$ direction, they are not plotted. Here, the  \emph{solid lines} stand for the
input field and the \emph{dashed lines} for the recovered one. It is clear that the
two fields are very similar and that the $z$ component is the best recovered
(see section \ref{sec:var_SNR}).
The \emph{bottom left panel} shows a map of $\abs{B}$ for a transverse section
after smoothing. The smoothing is made by convolving the field with a four
pixels full-width at half maximum (FWHM) gaussian. The green features
represent the input field and the reconstructed one is shown in the
superposed white contours.
The \emph{bottom right panel} shows the power spectra of the input field
(\emph{solid line}) and the recovered one (\emph{crosses}). 
Finally, figure \ref{fig:field_lines} represents the field lines of the input
field (\emph{top}) and the recovered one (\emph{bottom}).
These figures  show that, if the frequencies
are correctly sampled (see section \ref{sec:cond_los}), the  linear inverse problem (I)
recovers  qualitatively well the underlying field.  The local and global properties
of the field can be reconstructed provided that the linearization remains
valid which will be investigated in section \ref{sec:valid}.

It is of interest to study the conditioning of the linear problem for two
reasons (i) to understand the spatial spectral feature of the solution; in
particular the biases of the eigenvectors of the linearized problem which
induces anisotropy in the distribution of errors around the solution; (ii) to
constrain the best sampling strategy in order to recover $\V B$. Eventually it
will also have an impact on our ability to carry out the non linear
reconstruction.

The requirements to  set up a good conditioning of the global inverse problem can be formulated in steps.
First a necessary condition is to make a proper choice of the  (electromagnetic) frequency sampling, which can
be achieved by  looking at a smaller subproblem on a given LOS; however, this  optimal sampling 
does not warrant a  good global conditioning; we therefore investigate the quality of the global linear reconstruction by looking at different elements of the reconstruction covariance matrix in (spatial) frequency space.
In particular, we will show that the quality of the reconstruction is anisotropic and depends on the components of the field, $\V B$, which is confirmed by looking at the eigenvectors of the covariance matrix for a low dimensional problem.

\subsection{Conditioning of a line of sight and frequency sampling}
\label{sec:cond_los}

One can see easily that in the relation between polarization and magnetic
field (equation~(\ref{eq:defP})), each line of sight is independent of the
other. The link between them is provided by the solenoidal condition. In this
subsection we will not consider this condition and the matrix
$(\pder{\V{P}}{\V{B}})_{\V{B}_{0}}$ becomes block-diagonal. Moreover, the
three components can be separated leading to three different matrices,
$(\pder{\V{P}}{\V{B}_x})$, $(\pder{\V{P}}{\V{B}_y})$ and
$(\pder{\V{P}}{\V{B}_z})$. The field $\V{B}_0$ is taken constant and its
modulus set at $1\,\mu$G. In this case, all blocks are the same and the study
of the conditioning is reduced to the study of three $N_{\nu}\times N_{\V r}$
matrices with $N_{\nu}$ the number of frequencies and $N_{\V r}$ the number of
pixels in the $z$ direction.

Numerical investigations show that the conditioning of
$(\pder{\V{P}}{\V{B}_x})$ depends mainly on the
ratio $K\,h\,\Ne\,B_z/\nu^2$ leading to the conclusion that the conditioning is
dominated by the exponential term of equation (\ref{eq:defdPdBx}). It follows
that $(\pder{\V{P}}{\V{B}_y})$ has the same behavior as
$(\pder{\V{P}}{\V{B}_x})$ since the exponential terms are the same in both
equations (\ref{eq:defdPdBx}) and (\ref{eq:defdPdBy}), which is confirmed
numerically.

Recall that since  in this section the reference field is chosen constant, so is $B_z$;  therefore the
best sampling for the frequencies is to have $\nu^{-2}_{n} - \nu^{-2}_{n+1}$
constant, that is a constant step for the squared wavelength; hence:
$\lambda_n^2\equiv\lambda_0^2 + (n - 1)\,\Delta\lambda^2$ with
$n=1,\ldots,N_{\nu}$ the index of the frequency/wavelength.  So that the
complex exponential becomes
\begin{eqnarray}
    && \mathrm{e}^{2\,\mathrm{i}\,K\,\Ne\,B_z\,h\,m\,\lambda_n^2/c^2} \nonumber \\
    && = \mathrm{e}^{2\,\mathrm{i}\,K\,\Ne\,B_z\,h\,m\,\lambda_0^2/c^2}
    \mathrm{e}^{2\,\mathrm{i}\,K\,\Ne\,B_z\,h\,m\,(n -
    1)\,\Delta\lambda^2/c^2} 
\end{eqnarray}
with $m=1,\ldots,N_{\V r}$ the pixel index along the line of sight.  The value
of $\Delta\lambda^2$ must be chosen in such a way that the frequency dependent
complex exponentials are uniformly sampled on the complex circle.  Hence
$K\,\Ne\,B_z\,h\,N_{\V{r}}\,(n - 1)\,\Delta\lambda^2/c^2$ must be a multiple of
$\pi$ for any $n$.  With $L=h\,N_{\V r}$ the maximum probed depth and
taking the smallest multiple, this yields:
\begin{equation}
  \label{eq:dl2}
  \Delta\lambda^2 = \frac{\pi\,c^2}{K\,\Ne\,B_z\,L} \, .
\end{equation}
%
%
With this particular choice, the
matrices $(\pder{\V{P}}{\V{B}_x})$ and $(\pder{\V{P}}{\V{B}_y})$ take the
following form:
\begin{eqnarray}
  \nonumber
  (\pder{\V{P}}{\V{B}_{x/y}})_{n,m} = C_{x/y}e^{N_{\V
  r}i\beta}\left(\lambda_0^2+\frac{(n-1)\pi}{K\,\Ne B_zL}\right)^{\frac{\gamma-1}{4}} && \\
  \times \left(e^{-i\beta}e^{\frac{-2i\pi
      (n-1)}{N_{\V r}}}\right)^{N_{\V r}-m},\label{eq:dPdBxi}
\end{eqnarray}
where $\beta=2\,K\,\Ne\,B_z\,h\,\lambda_0^2$ and $C_{x/y}$ is a different constant in the
 $x$ and $y$ directions.
If the factor
$\left(\lambda_0^2+{(n-1)\pi}/({K\,\Ne B_zL})\right)^{\frac{\gamma-1}{4}}$ is set to 1,
the matrix is a unitary Vandermond matrix and its conditioning is 1 \citep{vandermonde}. 

Accounting for this factor impairs the conditioning but it stays close to unity.
The
elements of the last matrix, $(\pder{\V{P}}{\V{B}_z})$ are just geometrical
series of the elements of $(\pder{\V{P}}{\V{B}_x})$. Thus, they read:
\begin{eqnarray}
\nonumber 
\left({\partial
 \V{P}}/{\partial\V{B}_{z}}\right)_{n,m}&=&C_{z}e^{N_{\V
    r}i\beta}\left(\lambda_0^2+\frac{(n-1)\pi}{K\,\Ne B_zL}\right)^{\frac{\gamma+3}{4}} \times \\ && \hskip -3cm
\frac{\displaystyle 1-\exp({-i\beta(N_{\V r}+1-m)})\exp\left({\frac{-2i\pi}{N_{\V r}}(n-1)(N_{\V r}+1-m)}\right)}{\displaystyle 1-\exp\left({-i\beta}\exp({-\frac{2i\pi}{N_{\V r}}(n-1)})\right)}\,. \nonumber
\end{eqnarray}
where $C_z$ is yet another constant. 
At this stage, there is only one free parameter left, the first frequency $\lambda_0$.
The conditioning of $\left({\partial
 \V{P}}/{\partial\V{B}_{x/y}}\right)$ being always close to unity, the value of
$\lambda_0$
must be chosen in order to minimize the conditioning of $\left({\partial
 \V{P}}/{\partial\V{B}_{z}}\right)$.

\begin{figure}
\includegraphics[scale=0.55]{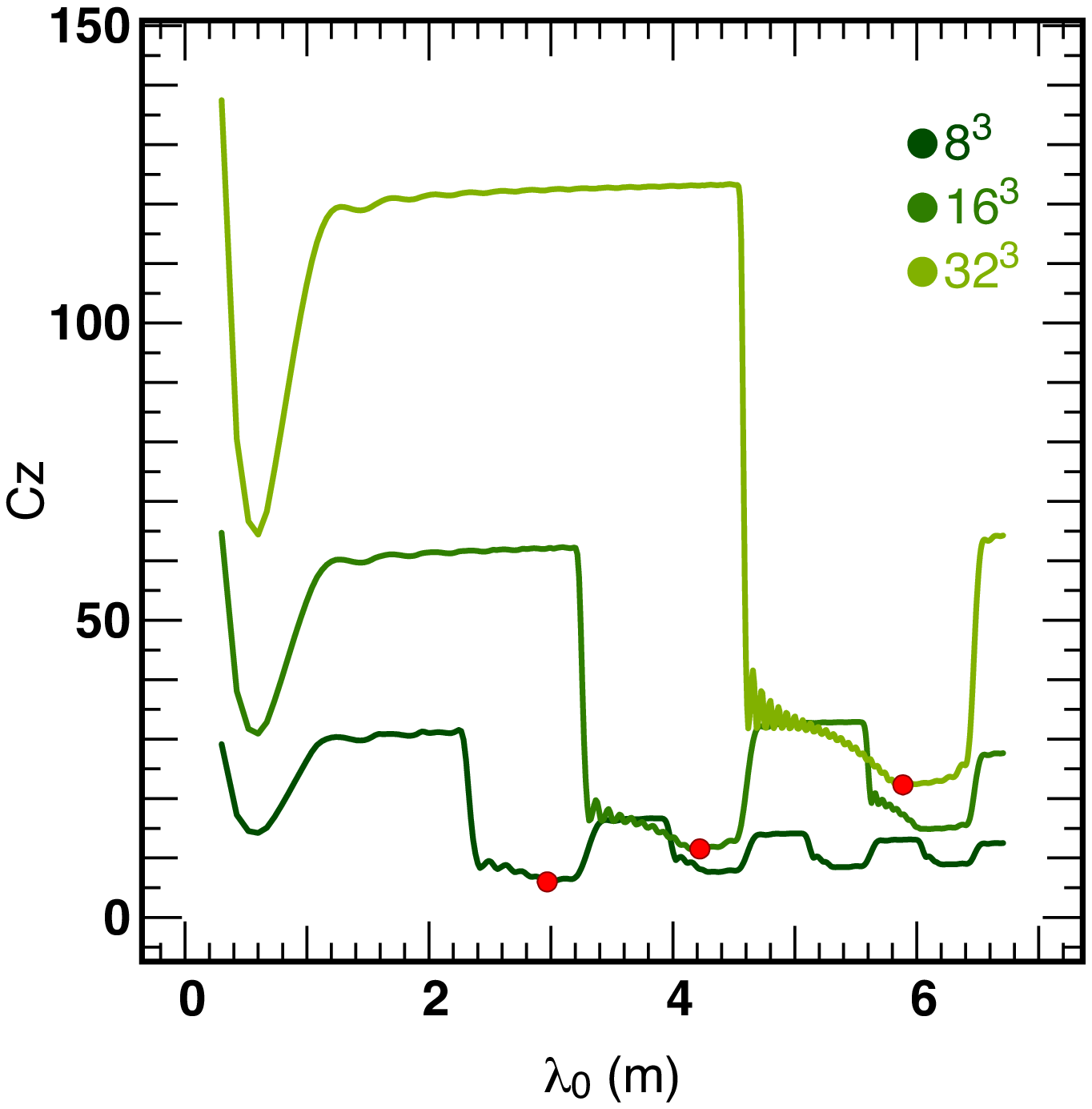}
\includegraphics[scale=0.55]{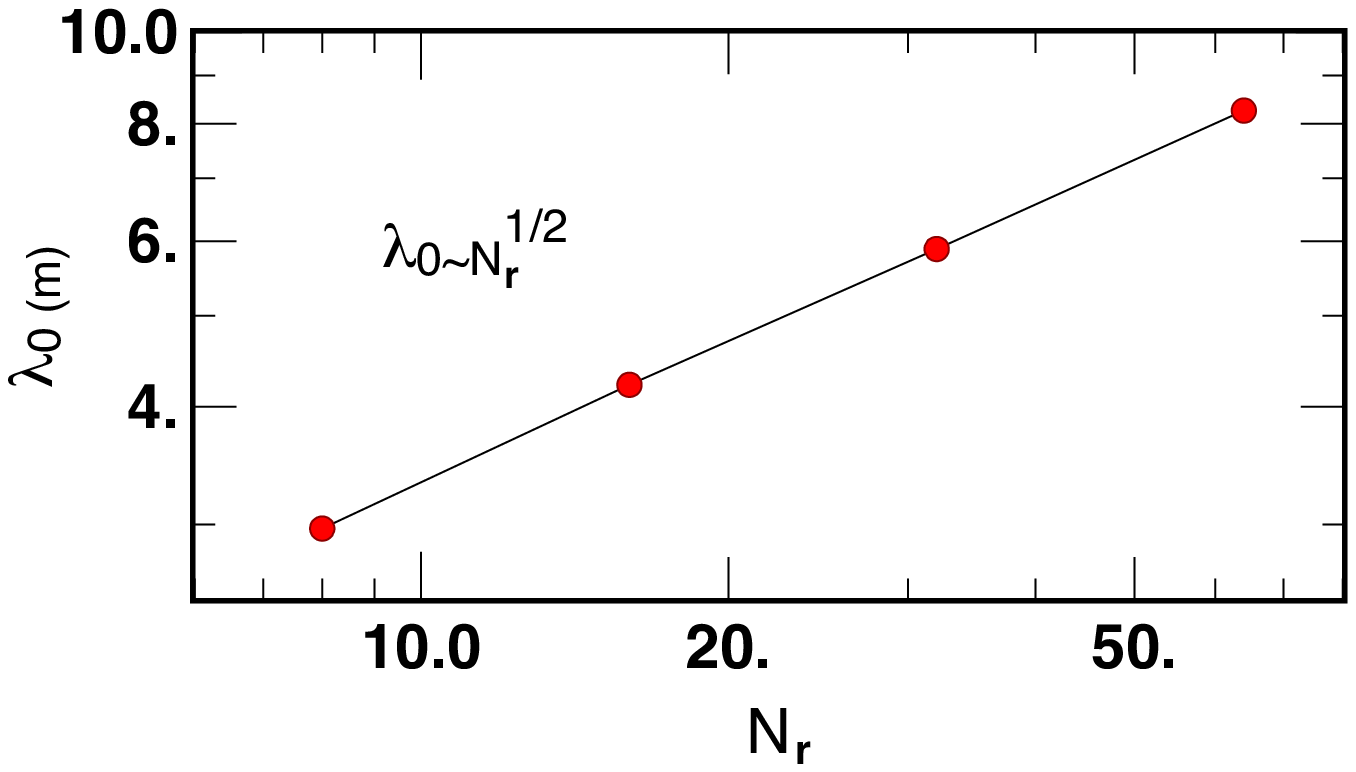}
\caption{ \emph{Top}: conditioning, $C_z$, of $\left({\partial
 \V{P}}/{\partial\V{B}_{z}}\right)$ as a function of $\lambda_0$ for different
grid sizes. The  red dots represent the best
conditionings. \emph{Bottom}: $\lambda_0$ giving the best conditioning as a
 function of the grid size, $N_\V{r}$. It appears that $\lambda_0\propto\sqrt{N_\V{r}}$.
 \label{fig:Lmin}}
\end{figure}

Figure \ref{fig:Lmin} (\emph{top panel}) represents the conditioning of $\left({\partial
 \V{P}}/{\partial\V{B}_{z}}\right)$ as a function of $\lambda_0$ for different
grid sizes. The curves are very similar in shape and the best
conditioning is represented by the red dots. In the \emph{bottom panel} the
wavelength providing the best conditioning for $\left({\partial
 \V{P}}/{\partial\V{B}_{z}}\right)$ is plotted as a function of the grid
size. It appears that $\lambda_0\propto\sqrt{N_{\V r}}$ and the precision on $\lambda_0$ is not really
important since the minimum of the curves are not really marked. These
particular choices of $\lambda_0$ give a conditioning of $1.29$ for $\left({\partial
 \V{P}}/{\partial\V{B}_{x/y}}\right)$, whatever grid size.

\subsection{Conditioning of $\Cmap$ and a
 posteriori variances }\label{sec:var_SNR}
\begin{figure*}
\includegraphics[scale=0.5]{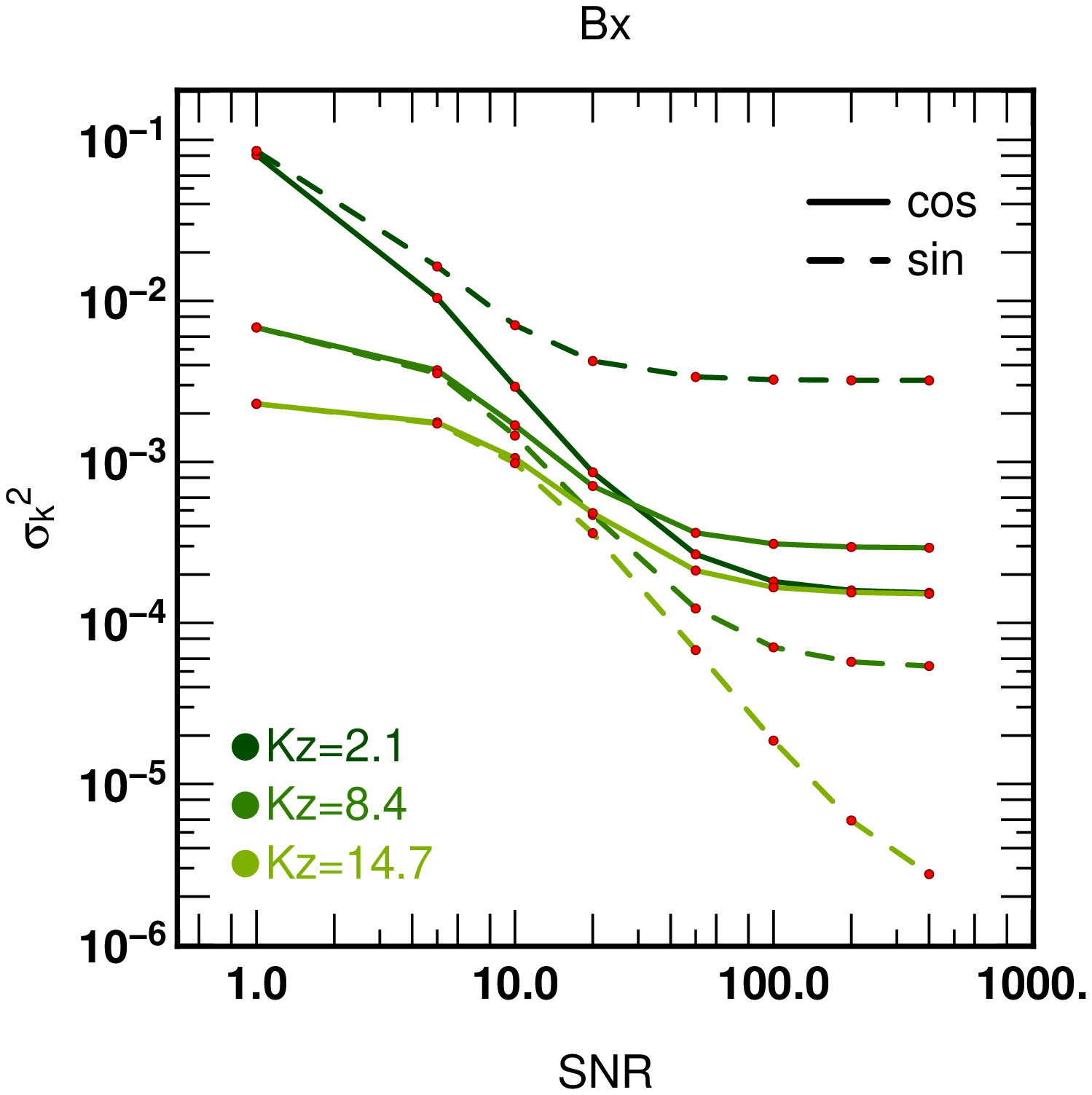}
\includegraphics[scale=0.5]{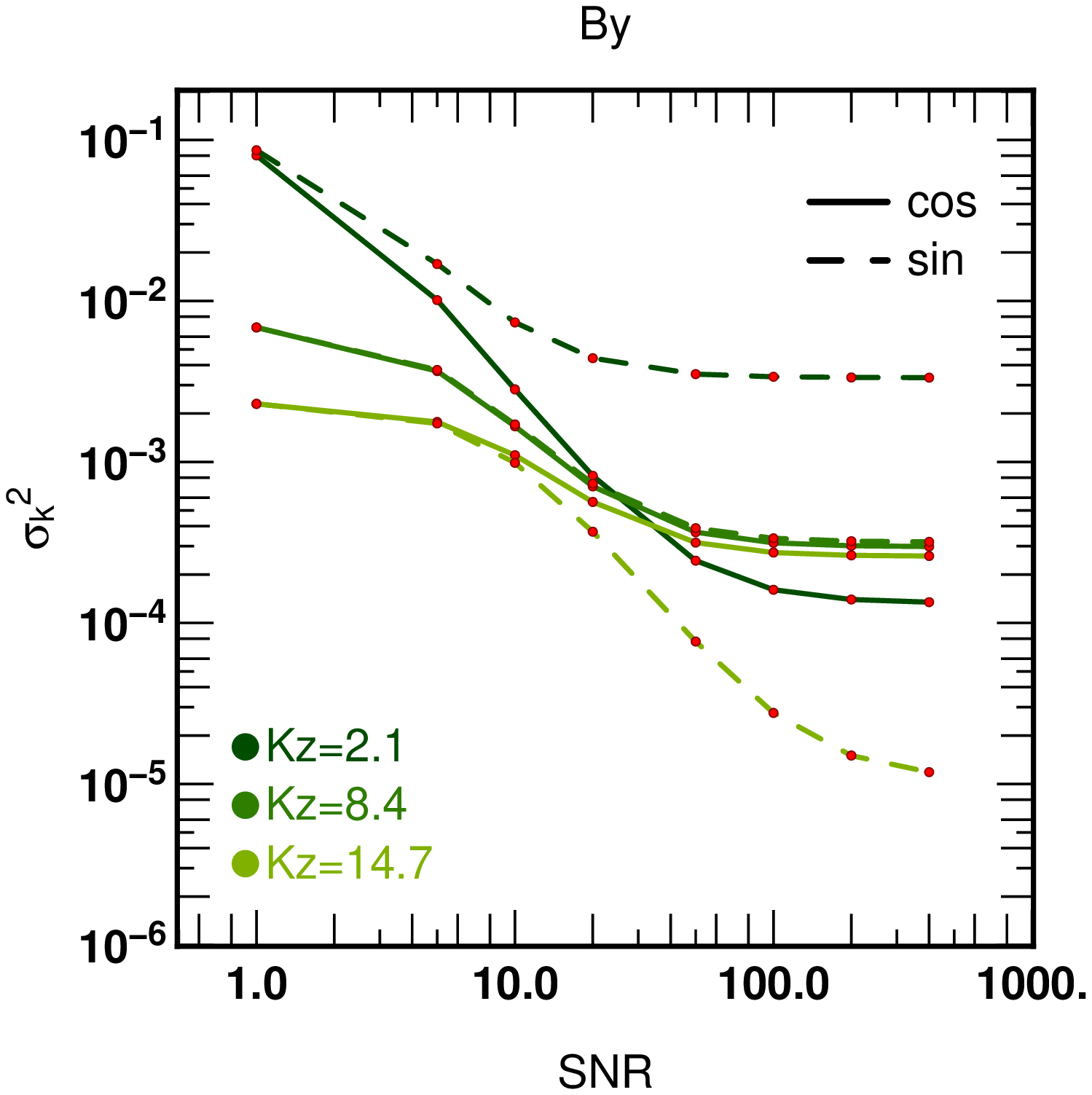}
\includegraphics[scale=0.5]{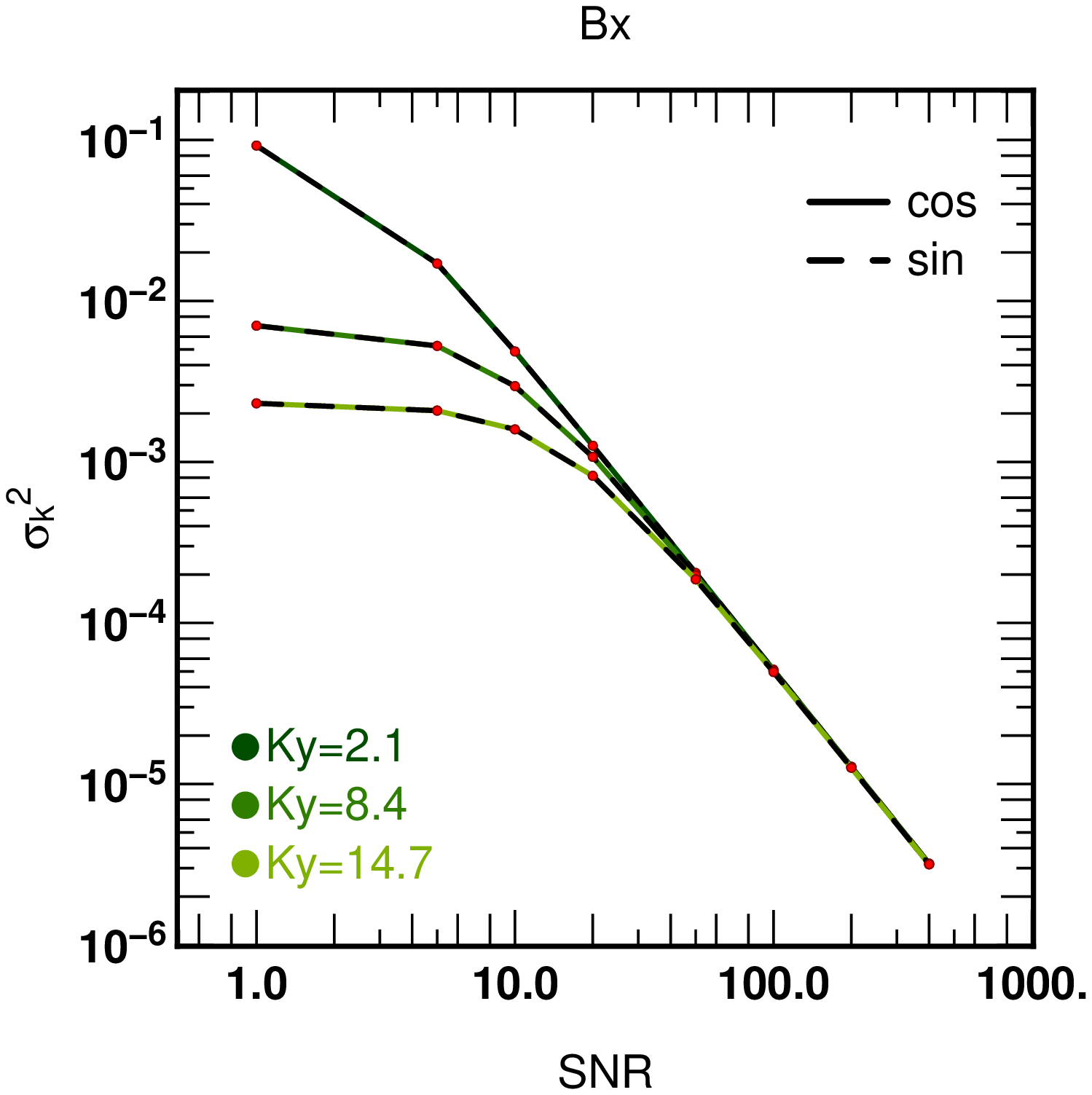}
\includegraphics[scale=0.5]{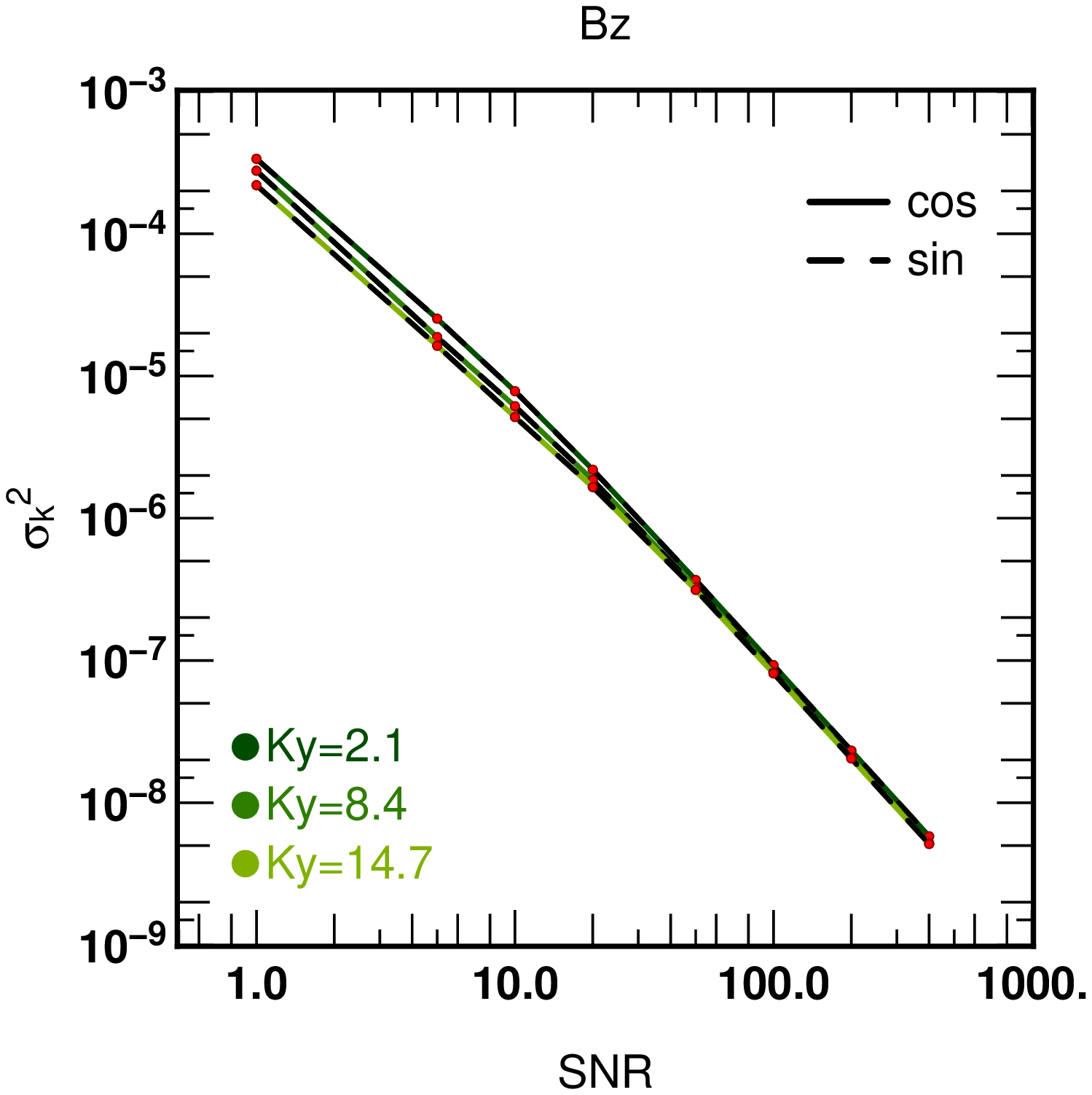}
\caption{A posteriori variance of different spatial frequencies $\V{k} \in
  (1,2,3)$ for the different components of the field in different directions
  (along a LOS or transverse to it) as a function of the SNR. The size of the
  box is $N_{\V r}=16$ and the number of frequency is $N_{\nu}=16$. The
  \emph{top panels} correspond to the variation of $\sigma_k^2$ for three
  different values of $k_z$ while the \emph{bottom panels} correspond to
  varying $k_y$.  The cosine mode (\emph{thick line}) and sine mode
  (\emph{dashed line}) are both shown. All variances decrease with increasing
  SNR as expected, although at different rate, see the main text.  Note the
  different amplitude in $\sigma^2_k$ for the \emph{bottom right panel} which
  shows that the $\V{B}_z$ component of the field is better recovered compared
  to the other components.  This reflects the anisotropy of the model $\V A $
  which induces anisotropic reconstruction errors.
 \label{fig:Var_SNR}}
\end{figure*}

\begin{figure*}
\includegraphics[scale=0.5]{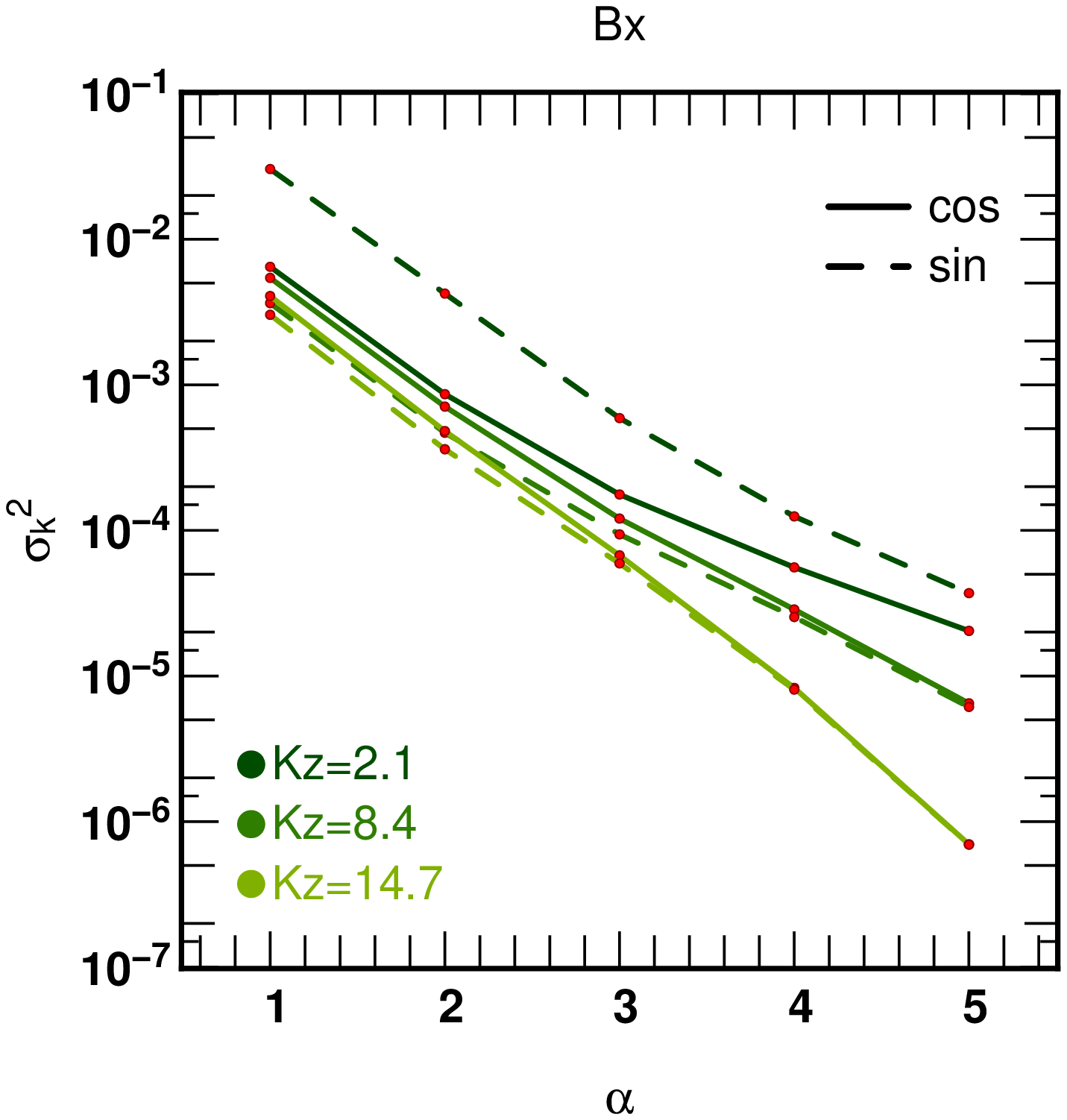}
\includegraphics[scale=0.5]{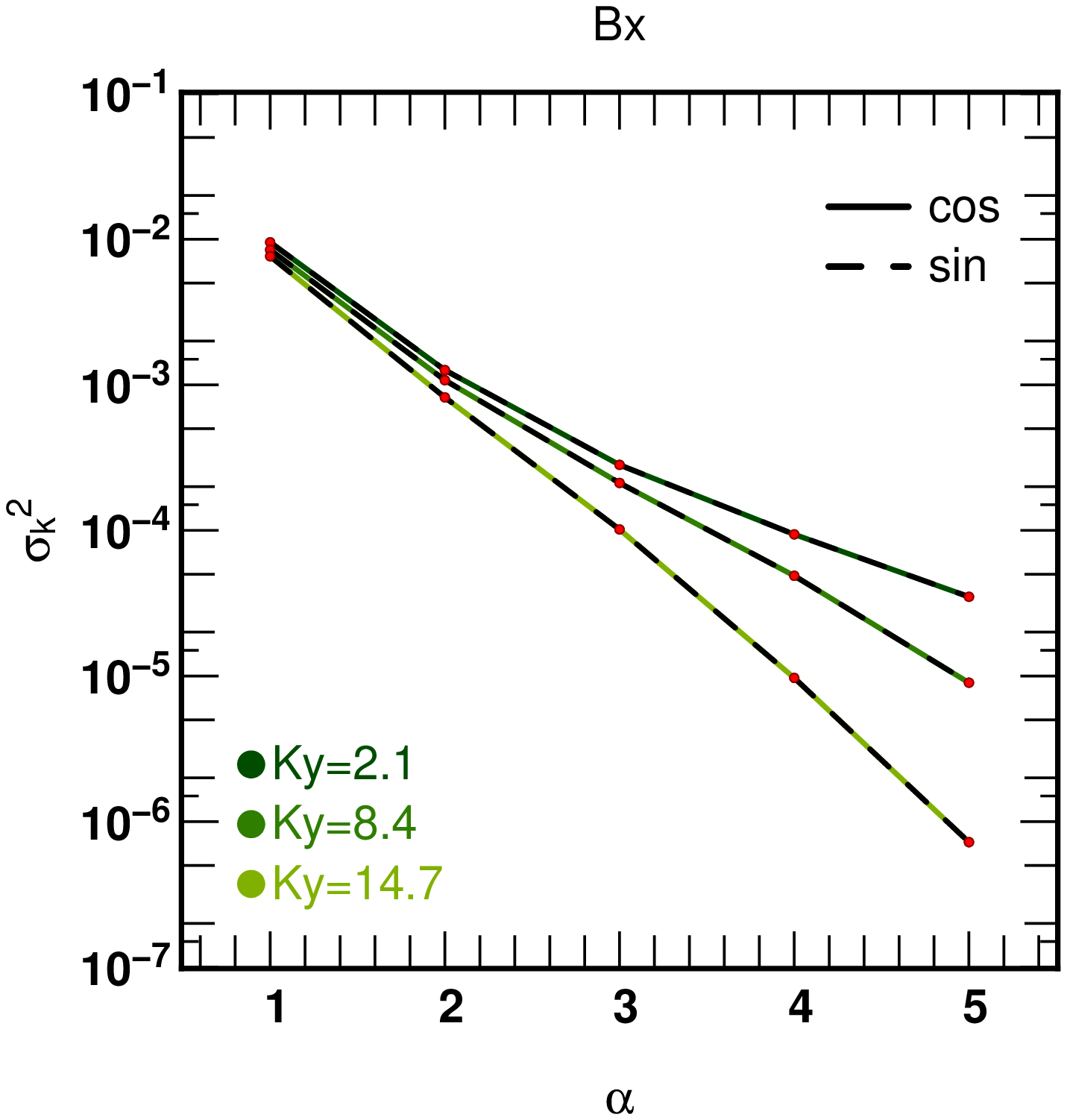}
\caption{ 
Same as Figure~\ref{fig:Var_SNR}  but as a function of  the spectral index,
$\alpha$, of $\delta\V{B}$ for a SNR$=20$.
As expected, the smoother the expected field, the larger $\alpha$,  the smaller the posterior variances.
 \label{fig:Var_gamma}}
\end{figure*}

Let us now investigate the a posteriori variances of different
spatial frequencies of the reconstructed field. This covariance matrix can be written as
\begin{equation}
\Cmap=(\M{A}^T\cdot \Cerr^{-1}\cdot\M{A}+\M{C}_\V{B}^{-1})^{-1},\label{eq:cmap}
\end{equation}
where $\M{A} \equiv\left({\partial\V{P}}/{\partial\V{B}_{0}}\right)\cdot\M{\Pi}$ with $\M{\Pi}$ the projector
that cancels the divergence (cf. equation~(\ref{eq:defproj})) and $\Cerr^{-1}$
and $\M{C}_\V{B}^{-1} \equiv \RegulWeight \, \V{F}^{-1}{\rm
  diag}(|k|^{\alpha})\V{F}$ are the a priori covariance matrices of the noise
and the signal respectively\footnote{Throughout this section (unless stated
  otherwise) we assume that $\alpha$ is given by minus the powerspectrum index
  of the sought magnetic field, and choose $\RegulWeight=1/{\rm P}(k=1)$,
  which corresponds to the minimum variance solution.}.  Here we seek
$\TFCmap$, the Fourier transform of $\Cmap$ as we want to understand the
relative error in the amplitude of the spatial modes of $\V B$.  Because of
the potential high dimensionality of our problem, the covariance matrix,
$\TFCmap$ is not computed directly. We chose instead to compute the
selected values by solving for $\V{\hat B}$ the following equation with a
conjugate gradient method
\citep[CGM,][]{shewchuk,Nocedal_Wright-2006-numerical_optimization}:
\begin{equation}
  \TFWmap\cdot \hat{\V{B}}=\hat{\V{B}}_{\rm ref}.
\end{equation}
Here, $\TFWmap=\TFCmap^{-1}$ and the solution, $\hat{\V{B}}$, found by the CGM
is
\begin{equation}
\hat{\V{B}}=\TFCmap \cdot \hat{\V{B}}_{\rm ref}.
\end{equation}
The reference field, $\hat{\V{B}}_{\rm ref}$, is equal to $1$ or $\pm i$ for
the chosen $\V{k}$ frequency and its opposite $-\V{k}$ in order to have a real
field, and $0$ elsewhere. The elements $\hat{\V{B}}_{\V{k}}$ and
$\hat{\V{B}}_{-\V{k}}$ of the solution are combinations of the covariance of
$\V{k}$ and $-\V{k}$ and the variance of $\V{k}$. It allows us to determine
the a posteriori variance of the chosen spatial frequency $\V{k}$.  To check
this method, the same variances were also computed by the iterative VMLM
method. One can check that:
\begin{equation}
  \langle(\hat{\V{B}}_{\rm in}-\hat{\V{B}}_{\rm out})\cdot(\hat{\V{B}}_{\rm in}-\hat{\V{B}}_{\rm out})^\dagger\rangle=\TFCmap,
\end {equation}
where $\dagger$ denotes conjugate transposition, $\hat{\V{B}}_{\rm in}$ and
$\hat{\V{B}}_{\rm out}$ stand respectively for the input field and the
reconstructed one in Fourier space.  As expected, the higher the number of
iterations, the closer the two estimates of the variance.
\\ Figure~\ref{fig:Var_SNR} represent the evolution of the a posteriori
variance of different spatial frequencies $\V{k}$ for the different components
of the field in different directions (along a LOS or transverse to it) as a
function of the SNR. The size of the box is $N_\V{r}=16$ and the number of
frequencies is $N_{\nu}=16$.  Figure \ref{fig:Var_gamma} shows the evolution
of the a posteriori variances of the same frequencies as of figure
\ref{fig:Var_SNR}, but as a function of the spectral index, $\alpha$, of the
sought field (for a SNR$=20$). As expected, the variance decreases as the
index increases.
In Figure~\ref{fig:Var_SNR} the SNR is defined as
\begin{equation}
{\rm SNR}={\rm RMS(data)}/\Serr,\label{eq:SNR}
\end{equation}
with $\Serr^2$ standing for the noise variance.
The results for the $B_y$ and $B_z$ fields in the $x$ direction are not plotted
because there are exactly the same as those in the $y$ direction.
First note that the variances, $\sigma^2_k$ for the $\V{B}_z$ component of the field are 
much smaller in amplitude relative to the other components.
For the $\V{B}_x$ and $\V{B}_y$ fields, at low SNR, the Wiener prior is
important in the
reconstruction, explaining the separation of the three curves corresponding to
three different scales.
 In Fourier space,
$\hat{\M{C}}_\V{B}=\RegulWeight^{-1}\,{\rm diag}(|k|^{-\alpha})$ with $\alpha$ the spectral index of the power
spectrum of the input field. If the regularization dominates,
$\TFCmap\sim|k|^{-\alpha}$,
which corresponds to the values on the figures when the SNR is low.

For the transverse frequencies (\emph{bottom panels}), the behaviour of the
variances is well understood. At low SNR, the Wiener prior dominates the
reconstruction for the $\V{B}_x$ and $\V{B}_y$ components but not for the
$\V{B}_z$ one. Increasing the SNR implies increasing the relative weight of the
data compared to the prior. So equation~(\ref{eq:cmap})
becomes
\begin{equation}
\Cmap\sim(\M{A}^T\cdot \Cerr^{-1} \cdot\M{A})^{-1},
\quad  \mbox{when SNR }\rightarrow \infty\,. \label{eq:CMAPasymptotic}
\end{equation}
If we assume a Gaussian white noise, $\Cerr=\Serr^2\,\M{I}$ with $\M{I}$
the identity matrix, equation~(\ref{eq:CMAPasymptotic}) becomes
\begin{equation}
\Cmap\sim\Serr^2(\M{A}^T\cdot \M{A})^{-1}\,,
\end{equation}
so $\Cmap\propto\Serr^2$ or given equation~(\ref{eq:SNR}),
$\Cmap\propto {\rm SNR}^{-2}$ which is the slope of these curves.
Finally, note that there is no symmetry breaking between the $x$ and $y$
directions and between the $x$ and $y$ components of the field or between the
sine and cosine modes in $\Cmap$.
\\
Now, consider the $x$ and $y$ components of the field along a LOS (\emph{top panels}). At low SNR,
the Wiener prior still dominate, providing the same value as in the transverse
direction. Then, the variance decreases as SNR${}^{-2}$ but reaches a threshold
and stagnate. It is clear on the figures that there is a symmetry breaking
between the $x$ and the $y$ components of the field and a separation between
the sine and cosine modes.
At first it may be surprising that the variances reach a threshold since the
frequencies have been chosen to provide the best possible conditioning for  
$\left({\partial \V{P}}/{\partial\V{B}_{0}}\right)$ along a LOS (see section
\ref{sec:cond_los}).
In fact this is a consequence of the solenoidal condition. Recall that 
for the global inverse problem, the relevant 
linear model is
$\M{ A}=\left({\partial \V{P}}/{\partial\V{B}_{0}}\right)\cdot\M{\Pi}$, where $\M{\Pi}$
is the projector given by equation~(\ref{eq:defproj}). This projector changes the matrix $\left({\partial \V{P}}/{\partial\V{B}_{0}}\right)$ and
adds off-diagonal terms to the block diagonal matrix considered in the
previous subsection. In effect, the solenoidal condition
degrades the   global conditioning  relative to the one LOS problem
(but recall that without it we have an ill posed problem).
 In turn this changes the eigen structure of $\TFCmap$ 
and therefore its projection in Fourier space.

Indeed,
 let us compute directly
the whole matrix $\TFCmap$  for a smaller, more tractable $N_{\V r}=8$  constant reference magnetic field
with $N_{\nu}=8$ frequencies sampled  following the procedure defined in  section
\ref{sec:cond_los}\footnote{As expected the curves of the
variance as a function of the SNR 
found previously
 are recovered exactly with this direct calculation.
}. 
 Figure \ref{fig:cond_cmap} shows the global conditioning of the 
covariance matrix $\Cmap$ as a function of the SNR.
One can see that the mixing of the LOS has a significant effect on conditioning,  even though the frequencies were chosen optimally. 
%
Figure \ref{fig:cond_cmap} also shows that at realistic SNR, the global
conditioning remains bounded  and could be improved, e.g. for the purpose of 
numerical convergence, by artificially increasing the
hyperparameter $\RegulWeight$.
Note finally that even though the global conditioning increases with the SNR,  the variances all decrease, as 
expected.

\subsection{Eigenspace analysis}\label{sec:spectral}
\begin{figure}
\includegraphics[scale=0.55]{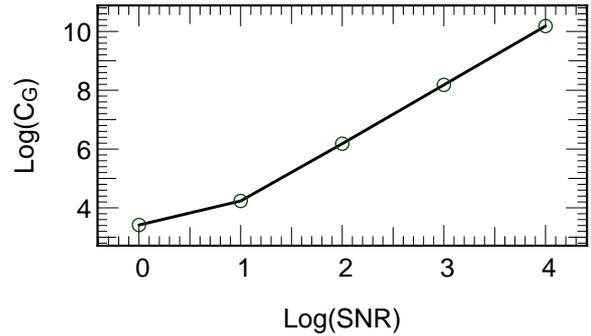}
\caption{Global conditioning of the (a posteriori)
covariance matrix $\Cmap$ as a function of the SNR.
The higher the signal to noise, to more difficult the inversion, but the
smaller the covariance a posteriori. The 3D matrix, $\M{
  A}=\left({\partial \V{P}}/{\partial\V{B}_{0}}\right)\cdot\M{\Pi}$ appears to
be  more poorly 
conditioned than its  1D counterpart  even though the sampling in 
electromagnetic frequency was the same as 
in section \ref{sec:cond_los}. It remains bounded and within reach of double precision calculation.
 \label{fig:cond_cmap}}
\end{figure}

\begin{figure*}
\includegraphics[scale=0.5]{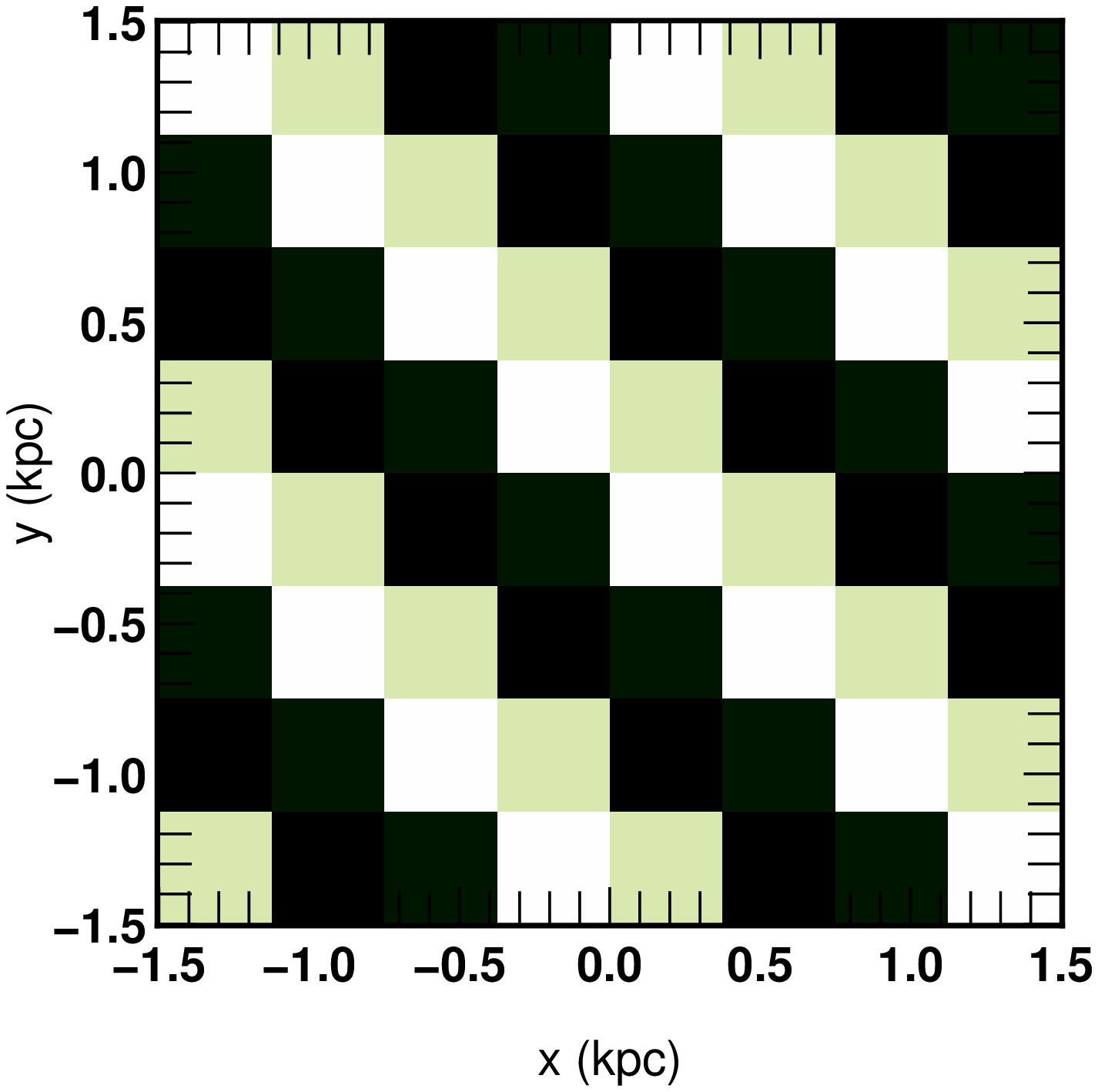}
\includegraphics[scale=0.5]{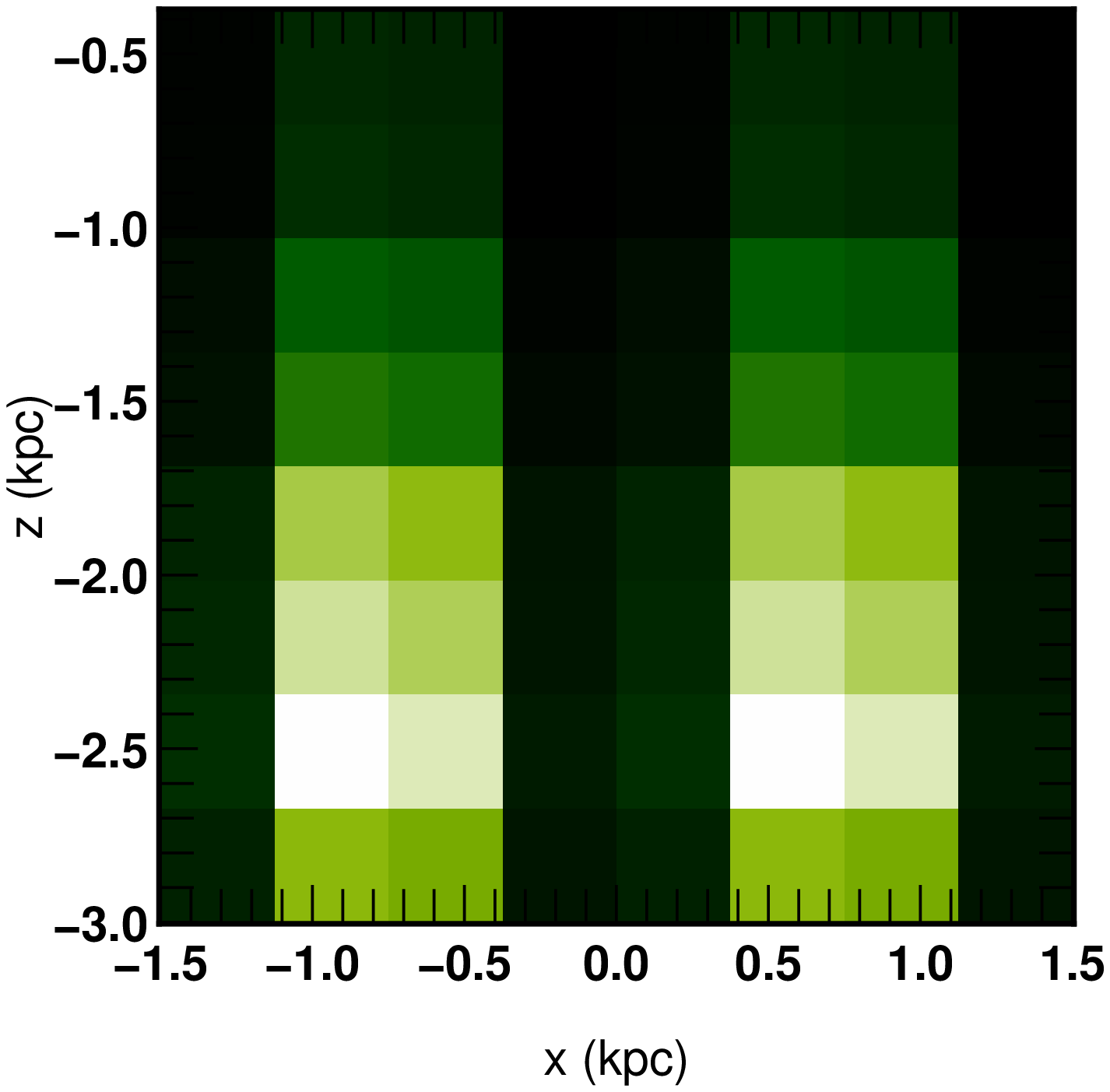}
\includegraphics[scale=0.5]{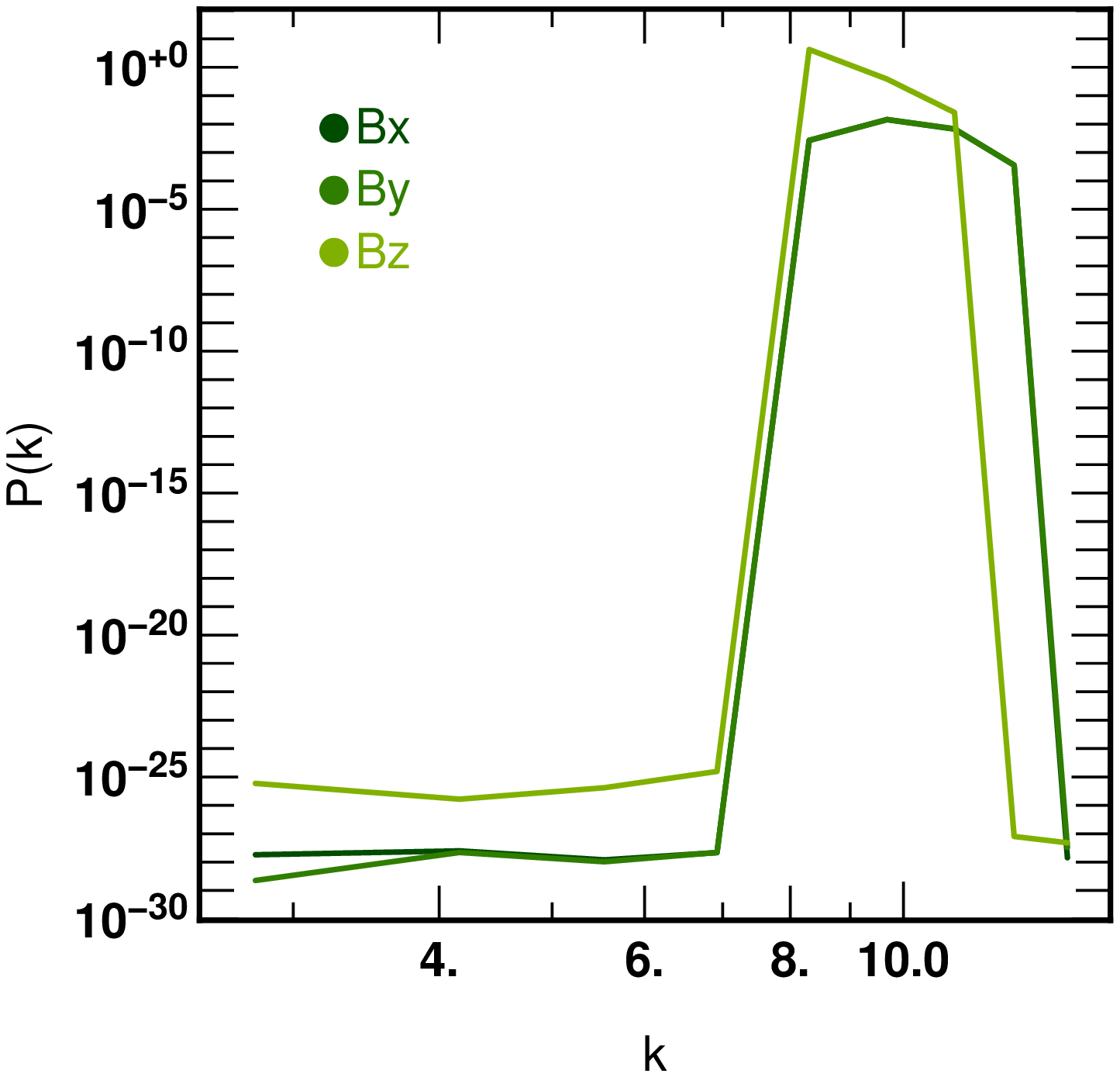}
\includegraphics[scale=0.5]{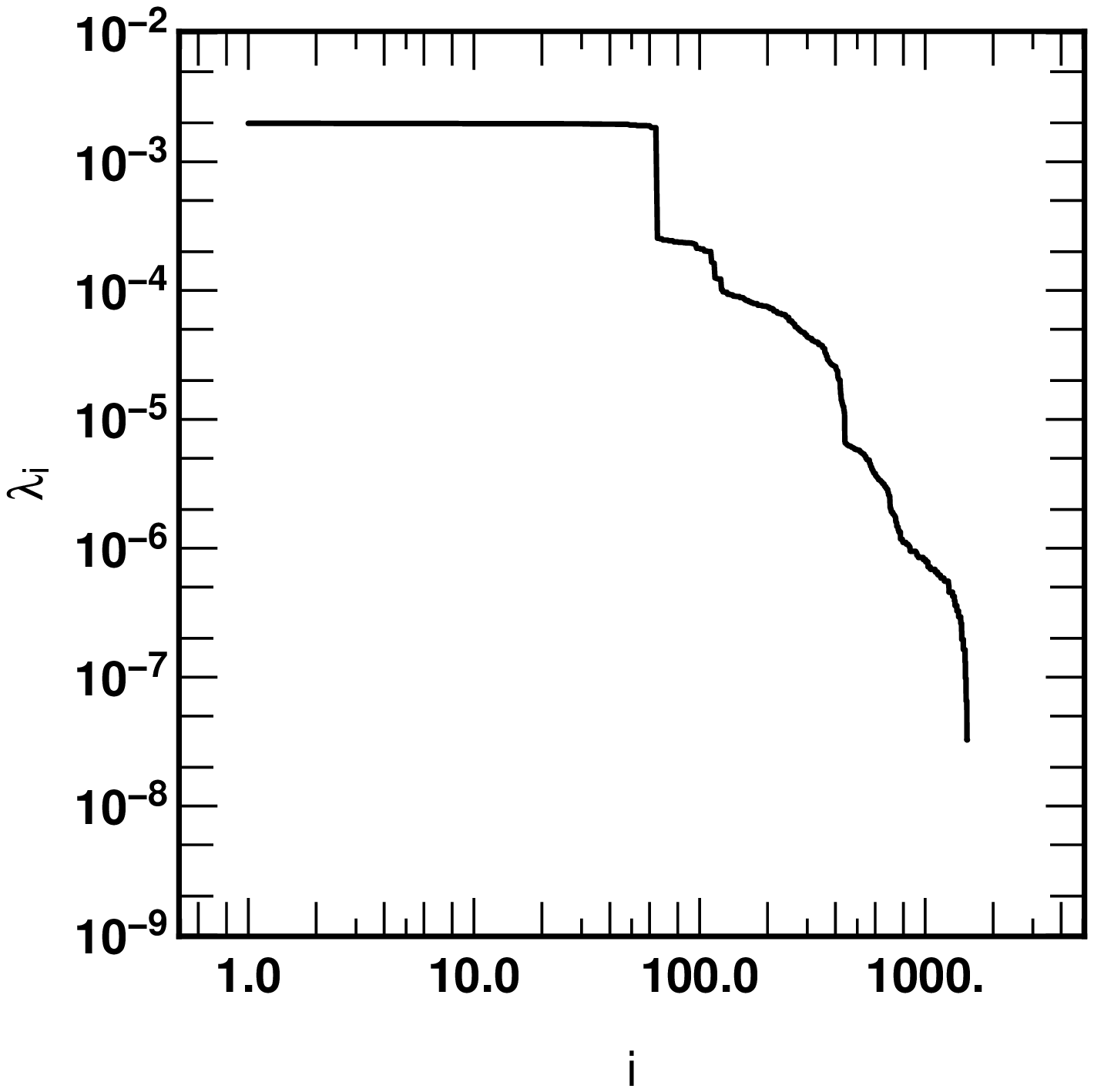}
\caption{ \emph{Top panels:} maps of the modulus of the field corresponding to
  the first eigenvector of $\Wmap$ in the $x\!-\!y$ (\emph{left}) and
  $x\!-\!z$ (\emph{right}) plans for a $8^3$ constant reference magnetic field with
  $N_{\nu}=8$ frequencies sampled as explained in section \ref{sec:cond_los} with a
  SNR$=20$.
The first eigenvector appears to be isotropic in $x$ and $y$ and anisotropic
in the $z$ direction.
\emph{Bottom left:} power spectra of the three components of this
  eigenvector. The anisotropy of the $z$ component is clearly visible and in
  good agreement with the results found in section \ref{sec:rec} (figure
  \ref{fig:64cube}) and \ref{sec:var_SNR} (figure \ref{fig:Var_SNR}).
\emph{Bottom right:} spectrum of the eigenvalues of $\Wmap$.
\label{fig:SVdec_Wmap}}
\end{figure*}

In order to understand the plateau on figure \ref{fig:Var_SNR},
let us also explicitly 
diagonalize $\Wmap$ for the smaller above-described
 $N_{\V r}=8$ problem with a SNR$=20$. The corresponding spectrum is plotted on figure
\ref{fig:SVdec_Wmap}, \emph{bottom right} panel. The  global conditioning of  $\Wmap$ is about
$10^{5}$  (consistently with what was shown on Figure \ref{fig:cond_cmap} for $\Cmap$), 
but note importantly that there is a cluster of eigenvalues followed by a
gap. This gap is consistent with the plateau seen on figure
\ref{fig:Var_SNR}. When increasing the SNR, 
one expects to filter out less and less eigen modes, and therefore  to access more and more
eigenvectors (corresponding to decreasing eigenvalues) in the
reconstruction. However, when reaching the gap, although the SNR increases, no
more eigenvalues are available for a while. The lower eigenvectors,
encoding informations on higher frequencies, are not within reach, and the a
posteriori variance of these frequencies stagnate, as  seen in figure
\ref{fig:Var_SNR}. If the   SNR increases further, these eigenvalues (and therefore their
associated eigenvectors) will be sampled, and we expect that the $\sigma^2_k$ variances will decrease
again\footnote{in other words, the plateau seen in the variance per mode in  the top panels 
reflects the fact that those modes have non zero contributions from the low signal to noise  eigen modes 
(i.e. eigen modes of $\M{C}_A^{-1/2} \cdot \M{C}_\V{B} \cdot \M{C}_A^{-1/2}$ with low eigen values,
where $\M{C}_A^{-1}\equiv \M{A}^T\cdot \Cerr^{-1} \cdot\M{A}$).
}.
The modulus of the first eigenvector (associated to the highest eigenvalue) is
plotted on the \emph{top panels} in the $x\!-\!y$ (\emph{left}) and $x\!-\!z$
(\emph{right}) planes. It is clear on these figures than the $x$ and $y$
directions are isotropic while the $z$ one is anisotropic for this
eigenvector. Moreover, the component of the power spectra in the \emph{bottom left
  panel} show that the $B_z$ component clearly differ from the
other two  components.

However, all of the main eigenvectors do not behave in the same way. Some of
them clearly break the symmetry between the $x$ and $y$ directions or/and between the $x$
and $y$ components leading to the differences in the curves of figure
\ref{fig:Var_SNR}. Finally note that the main eigenvectors are fairly high
frequencies fields. So, the a posteriori variances will be smaller for high
frequencies than for low ones, which is reflected by the \emph{top panels} of figure
\ref{fig:Var_SNR}.

\section{Validity of the linear approximation}\label{sec:valid}

\subsection {Linear and pseudo linear inversion}\label{sec:LPL}
\begin{figure}
\includegraphics[scale=0.5]{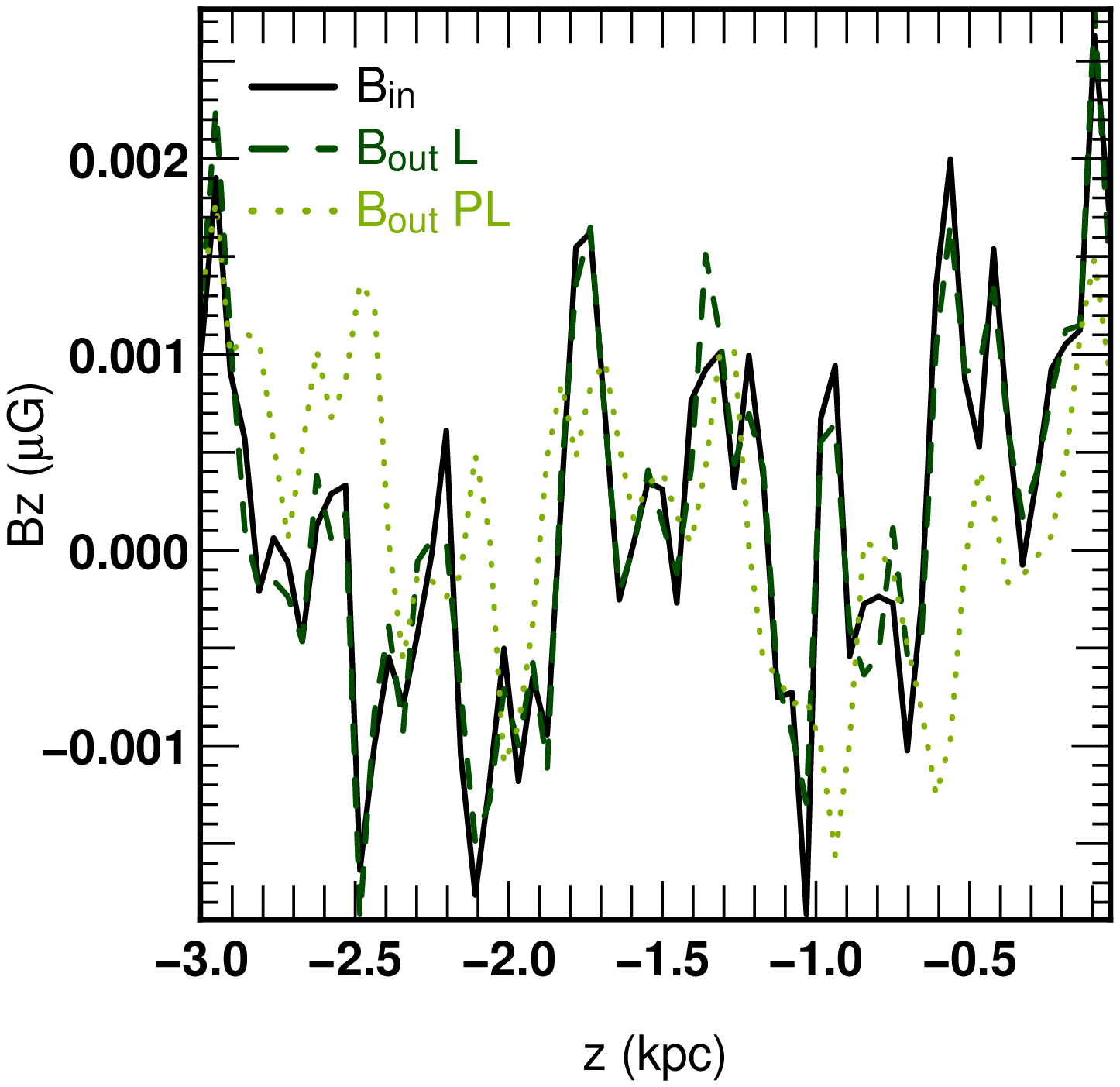}
\includegraphics[scale=0.5]{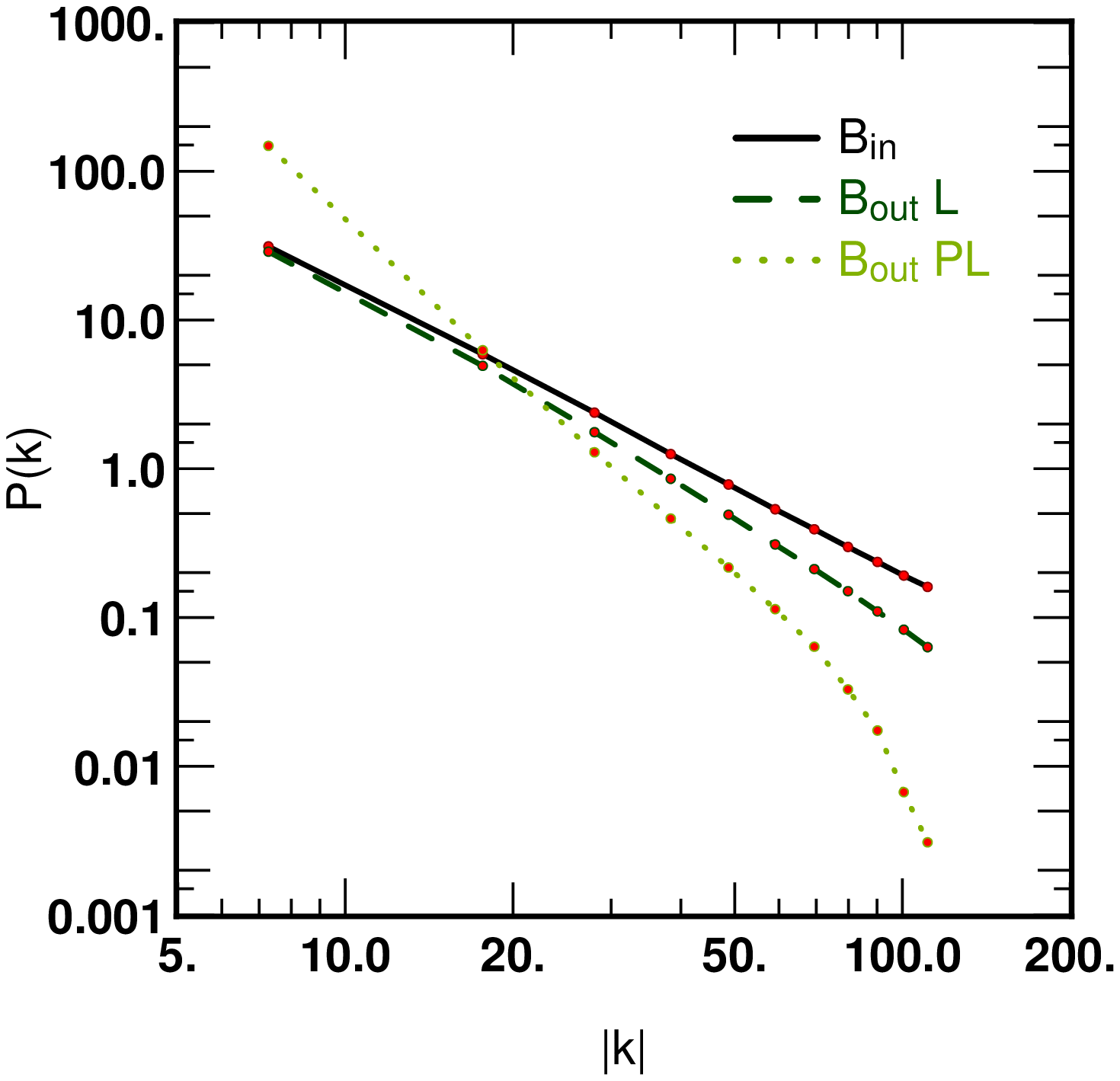}
\caption{ \emph{Top:} $\V{B}_z$ along a LOS for the input
  field (\emph{solid line}) and for the recovered fields with linear data, $\delta \V{P}_L$
  (\emph{dashed line}) and pseudo linear ones, $\delta \V{P}_{\rm PL}$ (\emph{dotted line}) (see section \ref{sec:linn}).
\emph{Bottom:} power spectra of these three fields. 
  Note that the power spectrum of the reconstructed field from the pseudo linear data set is steeper. 
  \label{fig:16cube}}
\end{figure}

 Let us
first carry out a linear inversion of the same pertubative field
$\delta\V{B}$, with RMS$(\delta\V{B})=10^{-3} \mu$G, while considering 
both  the linear (I) and  the pseudo linear (II) data sets (see section~\ref{sec:linn}). 
We work here on a $N_{\V r}=64$ grid, with
$N_{\nu}=64$ 
frequencies, a constant reference field of module 1 $\mu$G and SNR=$20$. Recall 
that for the
linear minimum variance solution,
 the hyperparameter $\RegulWeight=1/{\rm P}(k=1)$ (see section \ref{sec:penalty}), while for the the pseudo linear data set it may be
tuned. Figure~\ref{fig:16cube} \emph{top panel} shows the input $z$
component  for the input field (\emph{solid line}) along a given LOS and the output ones
(\emph{dotted line} for the linear data, $\delta \V{P}_L$ and \emph{dashed line} for the pseudo linear, $\delta \V{P}_{\rm PL}$) while
the \emph{bottom panel} shows the different power spectra. 
As  previously, the  field  recovered from linearized data sets  fits quite well
the input one. The recovered pseudo linear field, though somewhat 
different from the linear one,  remains fairly close to the original field. 
The corresponding powers pectra are also shown on Figure \ref{fig:16cube} and confirm that 
the recovered field in setting (II) is quantitatively redder.
\subsection{Second order residuals}\label{sec:ordre2}

\begin{figure}\label{sec:non-linear}
\includegraphics[scale=0.5]{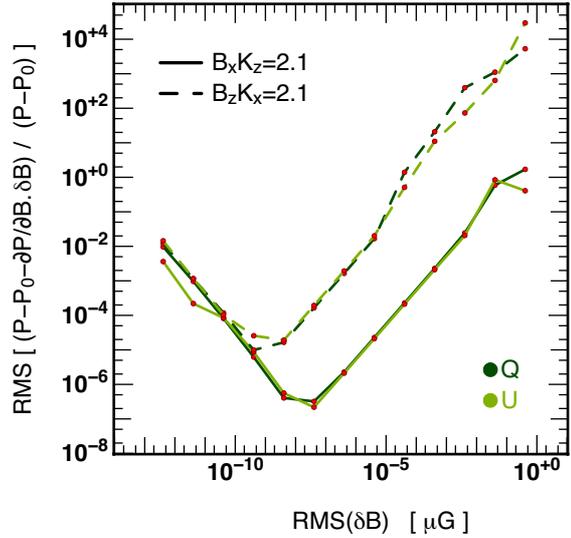}
\caption{ Average second order of the polarization divided by the first order
  as a function of RMS$(\delta\V{B})$. Here RMS$(\delta\V{B})$ is a single
  component and single mode field. Results are for a the lowest longitudinal
  mode for the $x$ component (solid lines) and the lowest transverse mode for
  the $z$ component (dashed lines). \emph{Dark curves} represent the real part $Q$
  of the polarization while \emph{light ones} are for the imaginary part $U$ (see
  equation (\ref{eq:PQU})).
 \label{fig:dBampli}}
\end{figure}

Let us now 
study the second order residuals to quantify the domain of validity of the linearization.
For this purpose, we subtract to the total polarization its zero and first order 
expansion to
obtain  ($\V{P}-\V{P}_0-\left({\partial
  \V{P}}/{\partial\V{B}}\right)_{\V{B}_{0}} \propto \delta\V{B}^2$) and
we divide this quantity by the first order term ($\V{P}-\V{P}_0 \propto \delta\V{B}$).
Figure~\ref{fig:dBampli} represents the average of this quantity as a function
of RMS$(\delta\V{B})$. Here the perturbation consist of  a single frequency and
single component field. The \emph{solid lines} represent the results obtained with a
$B_x$ component along the LOS at the lowest mode, while the \emph{dashed lines} correspond
to the lowest transverse mode of the $B_z$ component. The \emph{dark curves} represent the
real part, $Q$, of the polarization while \emph{light ones} stand for the imaginary part
$U$ (see equation (\ref{eq:PQU})).
At very low RMS$(\delta\V{B})$, numerical noise dominate but decreases as
the RMS increases. After reaching a minimum,  note  that the quantity
plotted increase as RMS$(\delta\V{B})$ since $\propto \delta\V{B}^2/
\delta\V{B}$ and thus $\propto \delta\V{B}$. As expected, the lower the RMS$(\delta\V{B})$, the better the
linear approximation and the better the reconstruction.
Note also the significant amplitude difference between the $\V{B}_z$ and $\V{B}_x$ components; 
we interpret this as a difference between the second derivatives of the field, which in turn, impairs
the accuracy of the linearization  for the $z$ component.
This should not be a limitation when carrying the non linear reconstruction using a method such as 
VMLM, as the amplitude  of the subsequent changes in the magnetic field will be scaled by the 
inverse second derivatives.  

\subsection{Towards the non linear problem}\label{sec:NL}

%
Up to now, we have only considered the situation where $\V {B}_0$ was assumed to be constant. What
happens to the conditioning when we add spatial frequencies to $\V {B}_0$
or/and over $\,\Ne$? It is
easy to see that adding transverse frequencies to the $x$ or the $y$ component of $B$
will not change the conditioning of a LOS. Indeed, according to equations
(\ref{eq:defdPdBx}) and (\ref{eq:dPdBxi}), only the constants $C_{x/y}$ are modified
and vary for each LOS, but remain constant along each of them,
which has no effect on conditioning. On the contrary, if the modulation is
along a LOS, $C_{x/y}$ is no longer constant, and varies for every pixel
along a LOS. However,  given that the conditioning is dominated by the exponential
terms in  the Vandermond approximation, it doesn't change dramatically. 
  Hence the choice of $\lambda_0$ and the sampling
frequency remain the same but the conditioning increases slightly;  it can reach $3$ for
$\left({\partial \V{P}}/{\partial\V{B}_{x/y}}\right)$ and  $40$ for
$\left({\partial \V{P}}/{\partial\V{B}_{z}}\right)$. 
\\
The situation is a priori more dramatic for  the $z$ component of the field or
for the electronic density $\,\Ne$. Indeed, 
the addition of a
transverse  modulation has significant consequences, as the value of $B_z$
(or/and $\,\Ne$) in
equation (\ref{eq:dl2}) becomes different for each LOS. Therefore, the value of
$\Delta\lambda^2$ should in principle be different for each LOS to conserve the best
conditioning.  In practice it is simplest to take
the average of $\V{B}_z$ (or/and $\,\Ne$) as a guess. 
However the conditioning per LOS increases signicantly and the
quality of the reconstruction should be affected. 
\\
However, it appears that the global conditioning of $\Cmap$ does not change
dramatically compared to the constant reference field value,  whatever the
frequency and the amplitude of the added
modulation. The  solenoidal condition appears to be very effective. In fact,
the  repetition of the spectral analysis carried in section \ref{sec:spectral}, shows that the main
difference will be in the gap seen on figure \ref{fig:SVdec_Wmap}. Adding
modulation on a constant field induces earlier, deeper gaps. 
At fixed SNR, the number of useful eigenvalues for the reconstruction decreases with the modulation.
 The inversion can still be carried, but will be more biased  by the lack of resolved eigenmodes.
 
\begin{figure*}
\includegraphics[scale=0.5]{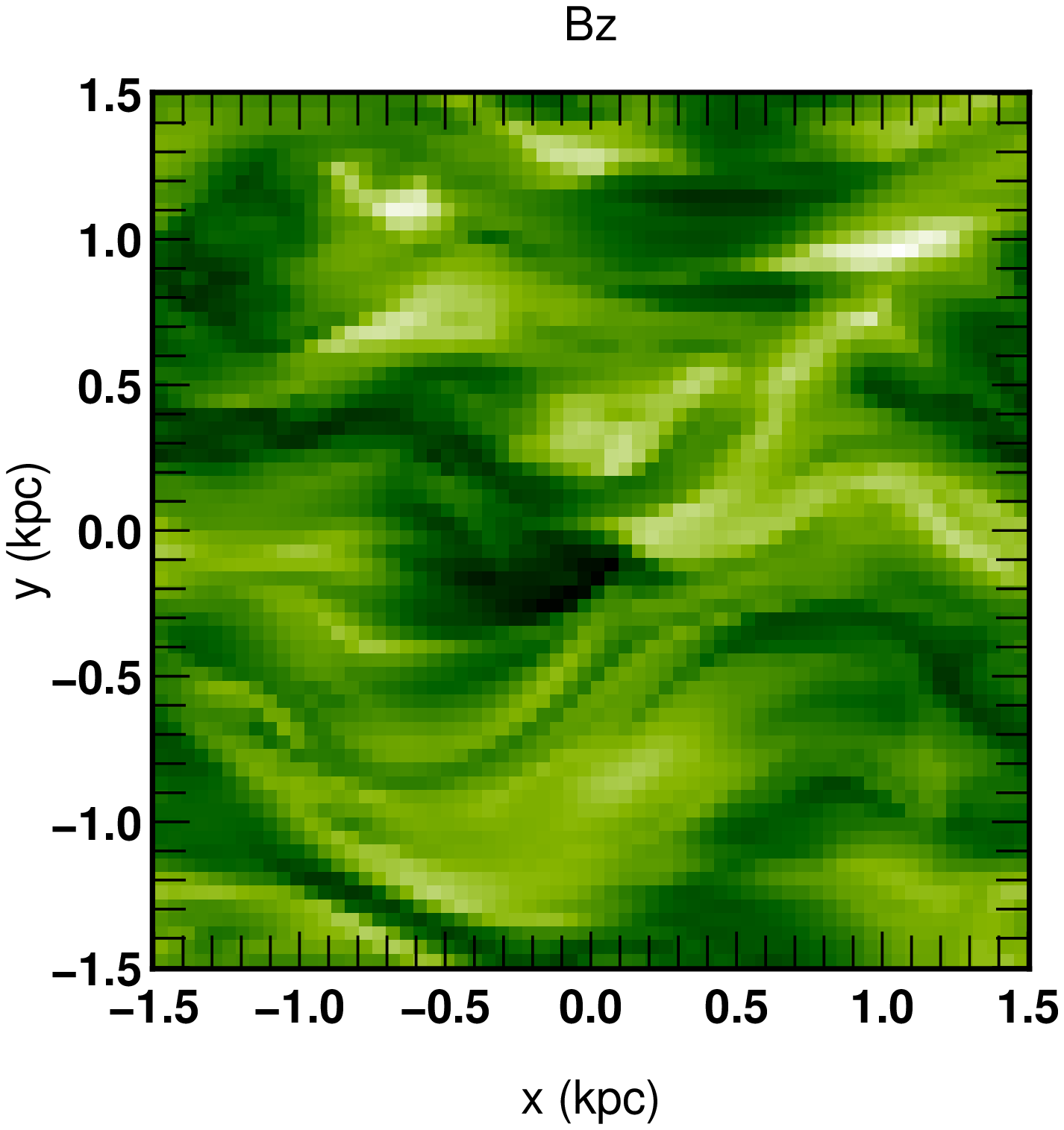}
\includegraphics[scale=0.5]{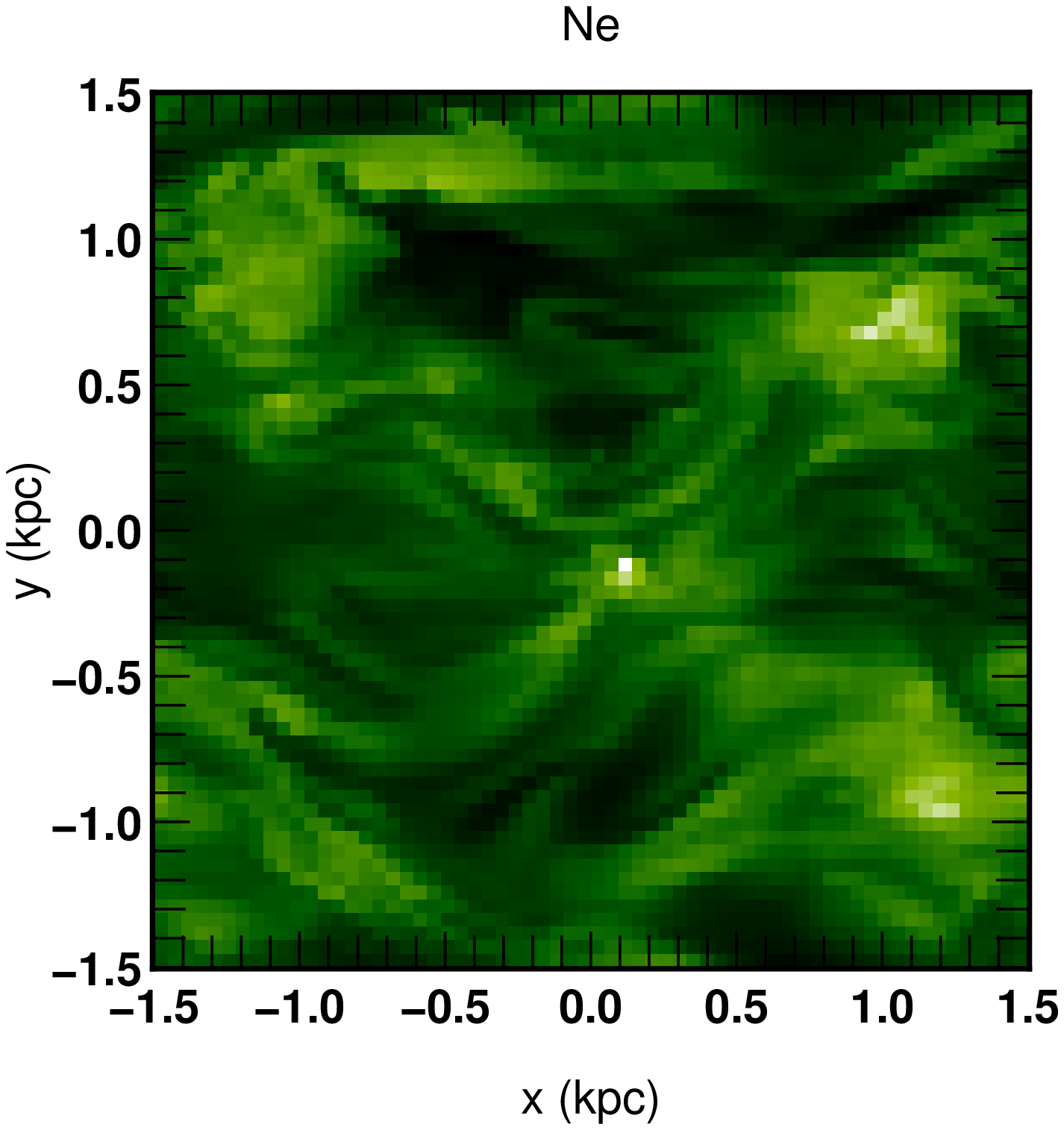}
\includegraphics[scale=0.5]{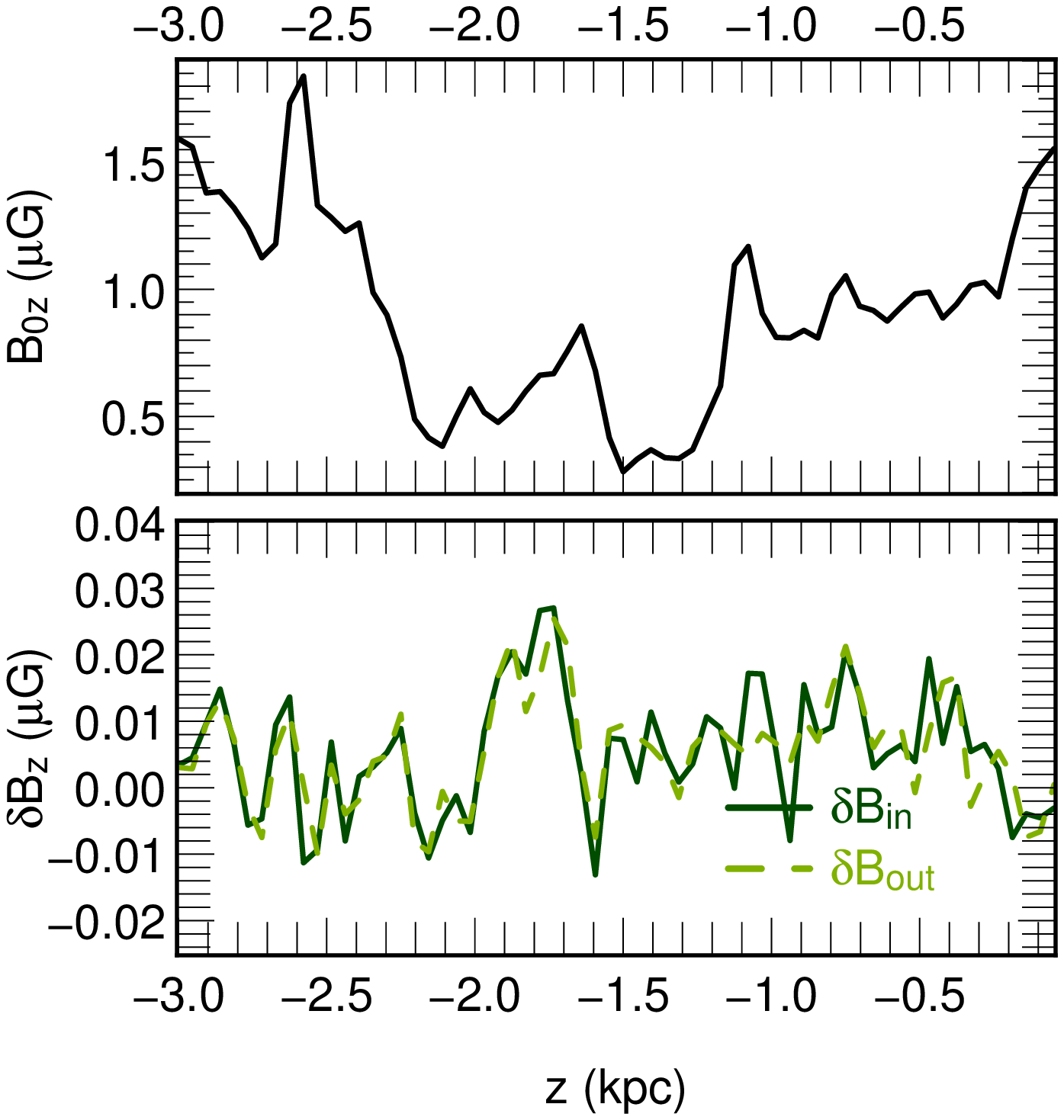}
\includegraphics[scale=0.5]{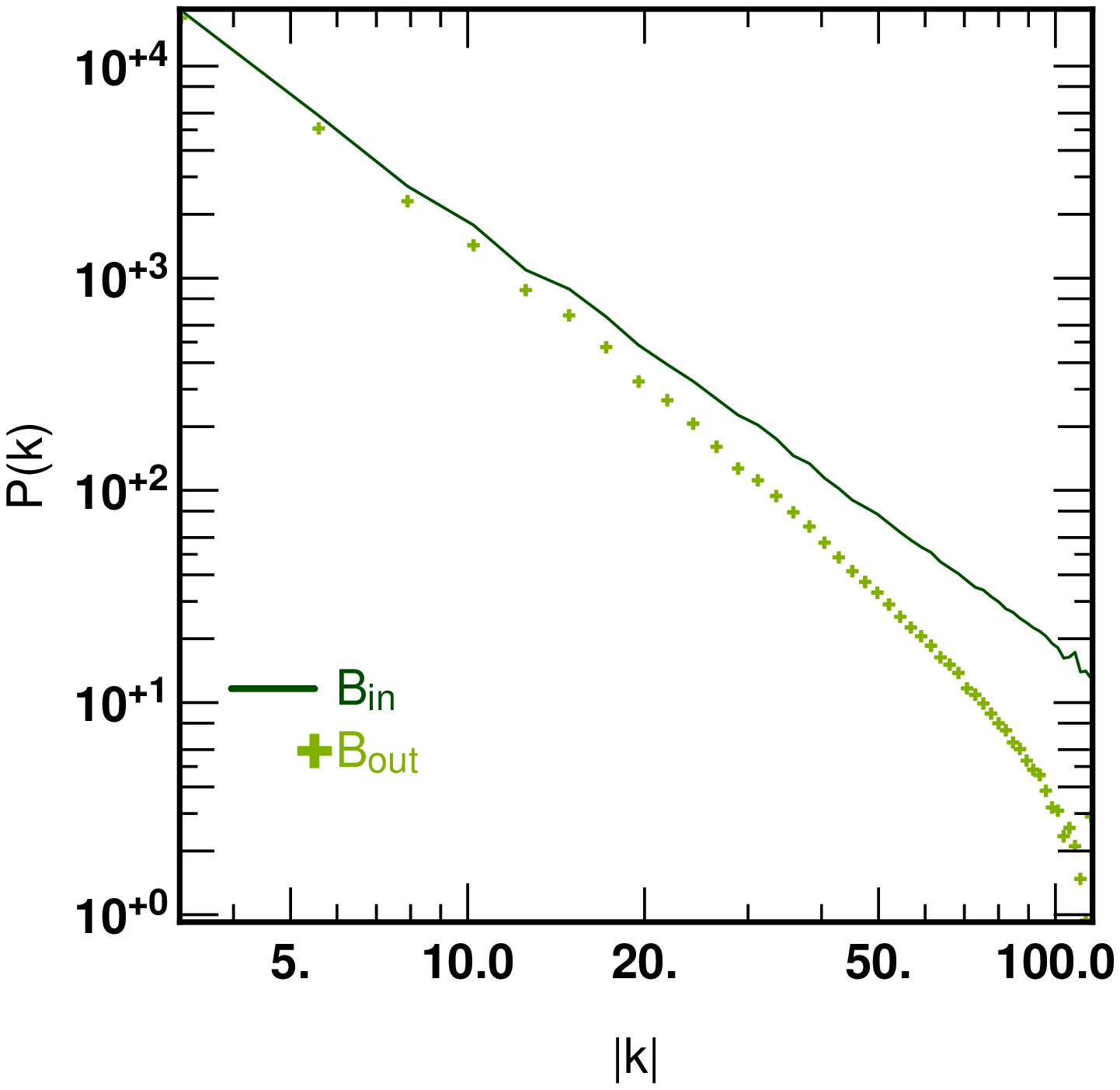}
\caption{ 
\emph{ left top panel}: map of a slice (of width $0.047$kpc) of input reference magnetic field, $\V{B}_0$; 
 \emph{right top panel}: map of the same slice but for the known electronic
 density $\,\Ne$;
  \emph{left bottom panel}: the input reference magnetic field, the input perturbation and the recovered one
  along a LOS. The perturbation field is a power-law fluctuation with a power
  spectrum of $\alpha=2$ and a relative amplitude of $10^{-2}$ from a virtual
  data set of SNR=20;
  \emph{right bottom panel}: input and recovered power spectra of the
  perturbation field.\label{fig:alex}}
\end{figure*}

 As a final illustration, figure~\ref{fig:alex} shows an implementation of the linear inversion on a more realistic
reference field, $\V{B}_0$ which is extracted from a 
magneto-hydrodynamical simulation \citep{refalex},
perturbed by a power-law fluctuation with a power spectrum of $\alpha=2$ and a
relative amplitude of $10^{-2}$ from a virtual data set of SNR=20. Note  that for this more realistic illustration the
electronic density $\,\Ne$ {\sl is not constant} but extracted from the same simulation.
Both the shape of the correction and its power-spectrum are
well recovered for this relative amplitude reflecting that although non constant model
and electronic density impair the conditionning, reconstructions remain possible.


\section{Conclusion and perspectives}\label{sec:Conclusion}

We investigated the problem of reconstructing the three-dimensional
spatial structure of the 
magnetic field of a given simulated patch of our Galaxy, using multi-frequency polarized maps
of the synchrotron emission at radio wavelengths.\\
When starting from a fair approximation of the magnetic field, we were
able to obtain a good estimate of the underlying field by using a linearized version of the
inverse problem considered, up to a $64^3$ grid size.  
The spectral analysis of the strictly linear problem (with a constant reference field,
and the simulated data obtained through a linearized model)
allowed us to specify the best sampling strategy in electromagnetic
frequency, 
and predict a spatially anisotropic distribution of
posterior errors.\\
The best sampling strategy is in equal $\Delta\lambda^2$; it follows from the 
shape of $\left({\partial \V{P}}/{\partial\V{B}_{0}}\right)$ along one LOS, which can be approximately recast into a 
unitary Vandermond matrix when this particular sampling is used.
The errors on the reconstructed  $B_x$ and $B_y$ components of the field
are shown to be larger than the error on the $B_z$ component.  
This anisotropy can be traced back to the shape of the posterior covariance, and ultimately of the linearized model 
which is highly anisotropic, as  only the $z$ component of the field induces Faraday rotation.

We considered in turn three more realistic cases: 
(i) a pseudo linear model (linear reconstruction of non-linearly simulated data),  (ii) a varying reference 
model $\V{B}_{0}$, and (iii) a varying reference model $\V{B}_{0}$ and a
(known) varying electronic density $\,\Ne$. We found that for these reconstructions, the 
 global conditioning of the minimum variance 
 solution remained tractable.
 Finally, we investigated the case where the reference field is given by the outcome of a magneto-hydrodynamical 
simulation, and is perturbed by an additional fluctuating component of known power spectrum. We showed that even in this
case the linear reconstruction quality is reasonable. This leads us to claim that a full non-linear reconstruction, based
on a Gauss-Newton sequence of linear sub-problems of varying reference field, should be achievable.\\

Possible extensions of this work, beyond the scope of this paper, involve
investigating systematically the degeneracies of the non-linear inversion. 
It would be worthwhile to construct specific  estimators for the (possibly anisotropic) local power spectrum of the field \citep[see e.g.][]{pogo}.
Finally, from a modelling point of view, 
one of the main limitations of the present method is that we had to assume known thermal and relativistic electronic
densities, in order to obtain a well posed inverse problem from synchrotron emission data alone. However, we could in principle
relax this assumption by adding extra data constraining the electronic
densities \citep[e.g. \Halpha data, see][]{haffner} or emission measures of pulsars, 
and attempt a joint reconstruction of the magnetic field and the electronic densities. 
Any prior  statistical information (e.g. extracted from MHD simulations) of possible correlation between $\V{B}$ and $\Ne$ could be used in this context. 
Another possibility would be to use the 
extra information given by the circular polarization of synchrotron emission (see Appendix~\ref{sec:Circular-Polarization}); 
this circular polarization, if negligible in the case
of low energy sources (like our Galaxy), is measurable in the case of relativistic radio sources \citep[see e.g.][]{Jones}, and opens
a way to constrain the electronic density together with the magnetic field structure of the source.

\subsection*{Acknowledgments}

\textit{{We thank Jean Heyvaerts, Martin Lemoine and
Guy Pelletier for fruitful comments on the early stages of this work.
Special thanks to Alex Lazarian for  providing us with his interstellar
magneto hydrodynamics simulations.
}}

\bibliographystyle{mn2e}
\bibliography{magbib}

\appendix

\section{The case $\gamma=3$}\label{sec:gamma3}
For $\gamma=3$, equation (\ref{eq:defP}) takes a particularly simple expression
\begin{eqnarray}
\nonumber
P=A\int_{-\infty}^{0}\nu^{-1}\,\Nr(z)(B_{x}^{2}(z)-B_{y}^{2}(z)+2iB_{x}(z)B_{y}(z))
&& \\
\times \exp\left(\frac{2iK}{\nu^{2}}\int_{z}^{0}(\,\Ne B_{z})(z'')dz''\right)\d z\,, &&
\end{eqnarray}
while equation (\ref{eq:defdPdBy}) simplifies to:
\begin{eqnarray}
\nonumber
\frac{\partial P(\V{x}_{\perp},\nu)}{\partial B_{x}({\bf
   \V{r}'})}&=&\delta_{\rm D}(\V{r}-{\bf \V{r}'})2Ah\nu^{-1}\,\Nr({\bf
 \V{r}}')(B_{x}+iB_{y})({\bf
 \V{r}}')
 \\ && \hskip -1.5cm
\times\exp\left(\frac{2iKh}{\nu^{2}}\sum_{z''}\theta_{\rm H}(z''-z')(\,\Ne B_{z})(x',y',z'')\right)\,. 
\end{eqnarray}
Note that for this value of $\gamma$ the two derivatives with respect to the transverse
magnetic field are thus related:
\begin{equation}
\frac{\partial P(\V{x}_{\perp},\nu)}{\partial B_{y}({\bf
   \V{r}'})}=i\frac{\partial P(\V{x}_{\perp},\nu)}{\partial B_{x}({\bf
   \V{r}'})}\,.
\end{equation}

\section{ Solenoidal fields with fixed power  spectrum.}\label{app:Bfields}

The generation of  solenoidal (divergence free) fields with fixed power
   spectra up to the Nyquist frequency is a  tricky problem. The field
must obey the three following conditions:
\begin{enumerate}
\item fixed power spectrum: $P(\V{k})\propto\V{k}^{-\alpha}$,
\item free divergence: $\V{\nabla}\cdot B\equiv0 \Leftrightarrow \bf{k}\cdot\hat{B}\equiv0$,
\item reality of the field: $\hat{B}_{\V{k}}=\hat{B}_{-\V{k}^{\star}}$.
\end{enumerate}
Given conditions (i) and (ii), the field is best generated in
Fourier space. Since the field is  multi periodic and we may write
\begin{eqnarray}
\hat{B}=\hat{B}_{\perp1}\V{e}_{\perp1}+\hat{B}_{\perp2}\V{e}_{\perp2}\,,
\end{eqnarray}
where $\V{e}_{\parallel}\equiv\V{k}/\abs{\V{k}}$, $\V{e}_{\perp1}\:{\rm
  and}\,\V{e}_{\perp2}$ form
a spherical basis in Fourier space, while $\hat{B}_{\perp,i}$,
i=1,2 are the projection over that basis of the Fourier componant
of the field. The vectors $\V{e}_{\perp1} \:{\rm
  and}\,\V{e}_{\perp2}$ are chosen in such a way that
  $\V{e}_{\V{k}\perp1/2}=-\V{e}_{-\V{k}\perp1/2}$. The spherical basis is
  direct for $\V{k}$ and indirect for $-\V{k}$. In this representation,
  conditions (ii) and (iii) become,
\begin{eqnarray}
\hat{B}_{\V{k}\perp1/2}=-\hat{B}_{-\V{k}\perp1/2}^{\star},\quad {\rm and} \quad
\hat{B}_{\V{k}\parallel}=0. \label{eq:cond}
\end{eqnarray}
So, the first step is to generate two complex fields $\hat{B}_{\perp1}$ and
$\hat{B}_{\perp2}$ with the sought power spectrum and then apply equation
(\ref{eq:cond}). 
\\
Next, consider the frequencies that have no conjugate, i.e. the
frequency $k_i=0$ (constant) and $k_i=N_y$ (Nyquist frequency) where the index
$i$ represents the Cartesian coordinates. Let us define $F_1$ as the set of
these two particular values, i.e. $F_1=[0,N_y]$, and $F_2$ the set of all the
other values, i.e. for a vector of dimension $N$, $F_2=[-(N/2-1),-(N/2-2),...,-1,1,...N/2-2,N/2-1]$.
When the three components of $\V{k}$ belong to $F_1$, the reality condition
of the field is merely $\Imag\hat{B}=0$.
After putting this imaginary part to $0$, the field can be projected into the
Cartesian basis.
\\
The difficulty arises when one or two components belong to $F_1$.
For example, consider the frequency $\V{k}=(k_x,k_y,k_z)$ with $k_x \in F_1$,
$k_y$ and $k_z\in F_2$. In this case, condition (iii) become
$\hat{B}_{\V{k}}=\hat{B}_{-\tilde{\V{k}}^{\star}}$ where
$\tilde{\V{k}}=(k_x,-k_y,-k_z)$ is the ``opposite'' of $\V{k}$. The problem is
that in this case, $\V{e}_{\V{k}\perp1/2}\neq-\V{e}_{-\tilde{\V{k}}\perp1/2}$
and the above discussed method  can no longer apply.
Fortunately, the combination of condition (ii) and
$\hat{B}_{\V{k}}=\hat{B}_{-\tilde{\V{k}}^{\star}}$ leads to the following set:
\begin{equation}
 \bf{k}\cdot\hat{B}\equiv0,\quad {\rm and }\quad
 \hat{B}_{\V{k}_x}=0.\label{eq:2D2}
\end{equation}
So, the trick is to put the faulty component to $0$ and to generate the
  other two as previously but in 2D space. Now, if $\V{k}_{{\rm 2D}}=(k_y,k_z)$, we
  generate $\hat{B}=\hat{B}_{\perp {\rm 2D}}\V{e}_{\perp \rm 2D}$, where
  $\V{e}_{\parallel \rm 2D} \equiv \V{k}_{\rm 2D}/\abs{\V{k}_{\rm 2D}}$ and $\V{e}_{\perp \rm 2D}$
  form a polar basis in Fourier space. As previously, the vectors
  $\V{e}_{\perp  \rm 2D}$ are chosen in such a way that
  $\V{e}_{\V{k}_{\rm 2D}\perp \rm 2D}=-\V{e}_{-\V{k_{\rm 2D}}\perp \rm 2D}$.
In this 2D representation, conditions (ii) and (iii) lead to:
\begin{equation}
\hat{B}_{\V{k}_{\rm 2D} \perp \rm 2D}=-\hat{B}_{-\V{k}_ {\rm 2D} \perp \rm 2D}^{\star},\label{eq:cond2}\\
\quad {\rm and}\quad
\hat{B}_{\V{k}_ {\rm 2D} \parallel \rm 2D}=0.
\end{equation}
Here we have only one degree of freedom left, thus, for these frequencies, we must generate one complex field $\hat{B}_{\perp \rm 2D}$ with the desired power
spectrum, and then apply equation (\ref{eq:cond2}). When $k_y$ or $k_z$
belongs to $F_1$, a similar procedure applies.
\\
In the last case, two component belong to $F_1$. For example,
$\V{k}=(k_x,k_y,k_z)$ with $k_x \in F_2$, $k_y$ and $k_z\in F_1$.
In this case, condition (iii) become
$\hat{B}_{\V{k}}=\hat{B}_{-\tilde{\V{k}}^{\star}}$ where
$\tilde{\V{k}}=(-k_x,k_y,k_z)$ is the ``opposite'' of $\V{k}$. Again, $\V{e}_{\V{k}\perp1/2}\neq-\V{e}_{-\tilde{\V{k}}\perp1/2}$ and the
combination of condition (ii) and
$\hat{B}_{\V{k}}=\hat{B}_{-\tilde{\V{k}}^{\star}}$ leads to 
equations (\ref{eq:2D2}). Consequently, the same
procedure follows for these frequencies.
After inverse Fourier transform, one can check that the field is real,
solenoidal and with the right power spectrum up to the Nyquist frequency.

\section{Circular Polarization}\label{sec:Circular-Polarization}

Since the rotating term depends on the density
field of thermal electrons $\Ne$ in the medium, we cannot
separate, with the Faraday rotation only, $\Ne$ from $B_{z}$.
One way to tackle this problem is to pick up the next coupling term
of the Stokes parameters in the (optically thin medium, strong rotativity
limit) assumption that describes our medium. This next term is a factor
of conversion between linear and circular polarization, that can be
considered together with the synchrotron emissivity of circular polarization \citep{Jones}.
Following the notations of \cite{Sazonov}, we write the transfer
equation of the polarization tensor $I_{\alpha\beta}$ as follows:
\begin{equation}
\frac{\ \d I_{\alpha\beta}(z)}{\d z}=E_{\alpha\beta}(z)-i(T_{\alpha\sigma}(z)\delta_{\beta\tau}-\delta_{\alpha\sigma}T_{\beta\tau}^{*}(z))I_{\sigma\tau}(z)\,,\label{eq:tensors}
\end{equation}
with $I_{\alpha\beta}=\left(\begin{array}{cc}
I+Q & U+iV\\
U-iV & I-Q\end{array}\right)$, and $E_{\alpha\beta}(z)$ is an emissivity term. In the assumption
of a thin, strongly rotating medium, we can retain only the rotating
terms (the Hermitian part) of $T_{\alpha\beta}$. Defining $T=\left(\begin{array}{cc}
h & q+if\\
q-if & -h\end{array}\right)$ we can show that the transfer equation can be reexpressed in terms
of the $(Q,V,U)$ {}``vector'' as:\begin{equation}
\frac{d}{dz}\left[\begin{array}{c}
Q\\
V\\
U\end{array}\right]=\left[\begin{array}{c}
E_{Q}\\
E_{V}\\
E_{U}\end{array}\right]-\left[\begin{array}{c}
h\\
f\\
q\end{array}\right]\times\left[\begin{array}{c}
Q\\
V\\
U\end{array}\right]\,.\label{eq:Stokes}\end{equation}
The fact that this differential equation involves multiplication by
a non-Abelian group element - in SO(3) - prevents us from writing a
formal solution to the equation in terms of exponentials. However,
since we are in the end working on a discretized mesh, we can still
write a formal solution to the discrete problem in terms of (finite)
sums of (finite) rotations products as we will see below. One important
point to notice, linked to the tensor nature of equation \ref{eq:tensors},
is the transformation law of these {}``vectors'' under rotation
of the coordinate axes in the plane perpendicular to the line of sight.
In this respect, the vector $(h,q,f)$ behaves the same way as the
vector $(Q,V,U)$, i.e. the $(Q,U)$and $(h,f)$ subvectors are rotated
by $2\psi$ when the coordinate axes are rotated by $\psi$. In the
case of a homogeneous medium, this allows \cite{Sazonov} and \cite{Jones}
to choose the coordinate axes used to measure $Q$ and $U$ so that the $V$
Stokes parameter couples only to $U$ (this is achieved when $q$
is set to $0$). In this reference frame, the projection of the (constant)
magnetic field is aligned with the second coordinate axis.

In the case of a fluctuating magnetic field, such a scheme is not
possible anymore, and we need to rotate the coupling coefficients
(best expressed in the reference frame given by the local projection
of the magnetic field) in a common, constant, reference frame. Thus,
in an inhomogeneous medium, the equation \ref{eq:Stokes} in the common
reference frame takes the form:\begin{equation}
\frac{d}{dz}\left[\begin{array}{c}
Q\\
V\\
U\end{array}\right]=\left[\begin{array}{c}
E_{Q}\cos(2\psi)\\
E_{V}\\
-E_{Q}\sin(2\psi)\end{array}\right]-\left[\begin{array}{c}
h\cos(2\psi)\\
f\\
-h\sin(2\psi)\end{array}\right]\times\left[\begin{array}{c}
Q\\
V\\
U\end{array}\right]\,,\label{eq:StokesQVU}\end{equation}
where $(Q,U,V)$ are measured in the common reference frame, and all
other quantities are defined in the frame of the local magnetic field.
In the applications we will consider in this paper, the rotation coefficients
are dominated by the contribution of cold (thermal) electrons of the
medium. In this context, $(h,f)$ take the following form \citep{Sazonov}:

\begin{displaymath}
  h =  -\frac{\Qe^{4}\,\Ne\,B_{\perp}^{2}}
              {4\,\pi^{2}\,\Me^{3}\,c^{3}\,\nu^{3}}
 \, , \quad
  f =  \frac{\Qe^{3}\,\Ne\,B_{\Vert}}
             {\pi\,\Me^{2}\,c^{2}\nu^{2}} \, .
\end{displaymath}
It is interesting to note that both the frequency dependence, and
the dependence on the magnetic field are different in the coupling
terms. We note that \ref{eq:StokesQVU} involves the multiplication
of the Stokes {}``vector'' by an element of a non-Abelian group
(SO(3)), which precludes finding a formal solution to this differential
equation. However, the linearity of the equation in the Stokes parameters,
allows us to write a formal solution in the discretized case in terms
of sums of products of rotations on the source terms. This equation
is very similar to the rigid body type equations encountered in mechanics,
with the (major) difference that it is linear. For simplicity, we
will consider here a first-order discretization of the problem (i.e.
we consider the different fields to be piecewise constant). The solution
to the homogeneous Stokes transfer equation can be written as:\[
\left[\begin{array}{c}
Q\\
V\\
U\end{array}\right](z)=\prod_{i=0}^{n_{z}-1}\exp(-\Delta zM_{i})\left[\begin{array}{c}
Q\\
V\\
U\end{array}\right](z=0)\,,\]
where $M_{i}$ is the skew-symmetric matrix corresponding to the vector
$(h_{i}\cos(2\psi_{i}),f_{i},-h_{i}\sin(2\psi_{i}))$. This discretized
solution ensures the exact conservation of the polarization degree
in the absence of internal sources. It corresponds to the simplest
possible case of integration of an equation on the SO(3) Lie Group
\citep[e.g.][]{Celledoni}. By linearity, we can find the discrete
solution to the transfer equation with sources (\ref{eq:StokesQVU}):\[
\left[\begin{array}{c}
Q\\
V\\
U\end{array}\right](z)=\sum_{i=0}^{n_{z}-1}\prod_{j=i}^{n_{z}-1}\exp(-\Delta zM_{j})\left[\begin{array}{c}
E_{Qi}\cos(2\psi_{i})\\
E_{Vi}\\
-E_{Qi}\sin(2\psi_{i})\end{array}\right]\,,\]
with $\psi=\arctan(B_{x}/B_{y})$ 
\footnote{Beware that this angle corresponds to $\pi-\psi$ in the notations
of section \ref{sec:emission}%
}. This expression generalizes equation~(\ref{eq:defPdisc}) and can be used to infer the Frechet derivatives of
the polarization field with respect to the magnetic field, as was
done in section \ref{sec:emission} from the integral solution.
The source terms in the frame attached to the local transverse magnetic
field read (\citet{Jones}):
\begin{eqnarray}
E_{Q}  =  C\,\Nr\,B_{\perp}^{\frac{\gamma+1}{2}}\nu^{-\frac{\gamma-1}{2}},\quad
E_{V} =  -D\,\Nr\,B_{\Vert}B_{\perp}^{\frac{\gamma}{2}}\nu^{-\frac{\gamma}{2}}
\label{eq:EQV}\end{eqnarray}
where $\Nr$ is the distribution of high-energy electrons in the medium, and C
and D are constants that depend on the energy distribution of relativistic
electrons. Note that in equation~(\ref{eq:EQV}), $n_r$ is weighted differently
in the expressions of $E_Q$ and $E_V$, hence we can in principle disentangle
$\V{B}$ from $n_r$. Here different assumptions can be made, namely assuming
either that $\Nr$ is related to the distribution of thermal electrons $\Ne$,
or that it is constant, or that it is related to the magnetic field pressure
locally (see \citet{beck2} for a discussion of the different
assumptions). Another possible path is to add external constraints on either
$\Ne$ (coming for instance from \Halpha observations \citep[see][]{haffner} or
from dispersion measurements of pulsars), or eventually on the relativistic
electron distribution $\Nr$ with diffuse gamma-ray measurements.

\end{document}